\documentstyle[12pt,epsf]{article}

\advance \topmargin by -\headheight
\advance \topmargin by -\headsep

\evensidemargin 
\oddsidemargin

\epsfverbosetrue
\def\figlabel#1{\xdef#1{\thefigure}}

\def\figalign#1#2#3#4#5#6{
\begin{figure}
\centerline{
\hbox to 2.5truein{\vtop{\hsize=2.5truein\epsfxsize=6cm
\centerline{\epsfbox{#1} }
\caption[]{#3}
\figlabel{#2} }}
\qquad\hbox to 2.5truein{\vtop{\hsize=2.5truein\epsfxsize=6cm
\centerline{\epsfbox{#4} }
\caption[]{#6}
\figlabel{#5} }} }
\end{figure} }

 \advance\hoffset by -3mm  
 \advance\voffset by  8mm  

\def\tr{{\hbox{\rm Tr}}}
\def\ex{{\hbox{\rm e}}}
\def\k{{\cal K}}

\newcommand{\CS}{{\scriptstyle {\rm CS}}}
\newcommand{\CSs}{{\scriptscriptstyle {\rm CS}}}
\newcommand{\ie}{{\it i.e.}}

\linethickness{.05 pt}
\newsavebox{\unoj}
\savebox{\unoj}(10,3){
\put(2,5) {\circle{10}}
\put(2,0){\line(0,0){10}
\put(0,4){${\scriptscriptstyle j}$}} }
\newcommand{\cunoj}{\mbox{\usebox{\unoj} \hskip 15pt  } } 

\newsavebox{\unodos}
\savebox{\unodos}(10,3){
\put(2,5) {\circle{10}}
\put(2,0){\line(0,0){10}
\put(0,4){${\scriptscriptstyle 2}$}} }
\newcommand{\cunodos}{\mbox{\usebox{\unodos} \hskip 15pt  } } 

\newsavebox{\unotres}
\savebox{\unotres}(10,3){
\put(2,5) {\circle{10}}
\put(2,0){\line(0,0){10}
\put(0,4){${\scriptscriptstyle 3}$}} }
\newcommand{\cunotres}{\mbox{\usebox{\unotres} \hskip 15pt  } } 

\newsavebox{\unocuatro}
\savebox{\unocuatro}(10,3){
\put(2,5) {\circle{10}}
\put(2,0){\line(0,0){10}
\put(0,4){${\scriptscriptstyle 4}$}} }
\newcommand{\cunocuatro}{\mbox{\usebox{\unocuatro} \hskip 15pt  } } 

\newsavebox{\dosi}
\savebox{\dosi}(10,3){
\put(2,5) {\circle{10}}
\put(0,0.3){\line(0,1){9.2}}
\put(4,0.3){\line(0,1){9.2} } }

\newsavebox{\dosii}
\savebox{\dosii}(10,3){
\put(2,5) {\circle{10}}
\put(-3,5){\line(3,0){10}}
\put(2,0){\line(0,5){10} } }
\newcommand{\cdosii}{\mbox{\usebox{\dosii} \hskip 15pt  } } 

\newsavebox{\dosiidosuno}
\savebox{\dosiidosuno}(10,3){
\put(2,5) {\circle{10}}
\put(-3,5){\line(3,0){10}}
\put(2,0){\line(0,5){10}
\put(0,7){${\scriptscriptstyle 2}$} } }
\newcommand{\cdosiidosuno}{\mbox{\usebox{\dosiidosuno} \hskip 15pt  } } 

\newsavebox{\dosiidosdos}
\savebox{\dosiidosdos}(10,3){
\put(2,5) {\circle{10}}
\put(-3,5){\line(3,0){10}}
\put(2,0){\line(0,5){10}
\put(0,7){${\scriptscriptstyle 2}$}
\put(-3,5){${\scriptscriptstyle 2}$} } }
\newcommand{\cdosiidosdos}{\mbox{\usebox{\dosiidosdos} \hskip 15pt  } } 

\newsavebox{\dosiitresuno}
\savebox{\dosiitresuno}(10,3){
\put(2,5) {\circle{10}}
\put(-3,5){\line(3,0){10}}
\put(2,0){\line(0,5){10}
\put(0,7){${\scriptscriptstyle 3}$} } }
\newcommand{\cdosiitresuno}{\mbox{\usebox{\dosiitresuno} \hskip 15pt  } } 

\newsavebox{\tresiii}
\savebox{\tresiii}(10,3){
\put(2,5) {\circle{10}}
\put(0,0.3){\line(0,1){9.2}}
\put(-3,5){\line(3,0){10}}
\put(4,0.4){\line(0,1){9.1} } }
\newcommand{\ctresiii}{\mbox{\usebox{\tresiii} \hskip 15pt  }
} 

\newsavebox{\tresiv}
\savebox{\tresiv}(10,3){
\put(2,5) {\circle{10}}
\put(-0.25,0.5){\line(1,2){4.45}}
\put(-3,5){\line(3,0){10}}
\put(4.25,0.6){\line(-1,2){4.45} } }
\newcommand{\ctresiv}{\mbox{\usebox{\tresiv} \hskip 15pt  }
} 

\newsavebox{\tresivdos}
\savebox{\tresivdos}(10,3){
\put(2,5) {\circle{10}}
\put(-0.25,0.5){\line(1,2){4.45}}
\put(-3,5){\line(3,0){10}}
\put(4.25,0.6){\line(-1,2){4.45}
\put(-4,4.5){${\scriptscriptstyle 2}$} } }
\newcommand{\ctresivdos}{\mbox{\usebox{\tresivdos} \hskip 15pt  }
} 

\newsavebox{\tresiiidos}
\savebox{\tresiiidos}(10,3){
\put(2,5) {\circle{10}}
\put(0,0.3){\line(0,1){9.2}}
\put(-3,5){\line(3,0){10}}
\put(4,0.4){\line(0,1){9.1}
\put(-4,6){${\scriptscriptstyle 2}$} } }
\newcommand{\ctresiiidos}{\mbox{\usebox{\tresiiidos} \hskip 15pt  }
} 

\newsavebox{\tresiiidosbis}
\savebox{\tresiiidosbis}(10,3){
\put(2,5) {\circle{10}}
\put(0,0.3){\line(0,1){9.2}}
\put(-3,5){\line(3,0){10}}
\put(4,0.4){\line(0,1){9.1}
\put(-2.5,5){${\scriptscriptstyle 2}$} } }
\newcommand{\ctresiiidosbis}{\mbox{\usebox{\tresiiidosbis} \hskip 15pt  }
} 

\newsavebox{\cuavi}
\savebox{\cuavi}(10,3){
\put(2,5) {\circle{10}}
\put(-1,0.8){\line(0,1){8.2}}
\put(2,0){\line(0,5){10} }
\put(5,1){\line(0,1){7.8} }
\put(-3,5){\line(3,0){10} }}
\newcommand{\ccuavi}{\mbox{\usebox{\cuavi} \hskip 15pt  } } 

\newsavebox{\cuavii}
\savebox{\cuavii}(10,3){
\put(2,5) {\circle{10}}
\put(-1,0.8){\line(0,1){8.1}}
\put(-2.7,3.1){\line(5,6){5.6} }
\put(6.5,3.1){\line(-5,6){5.6} }
\put(5,1){\line(0,1){7.8} } }
\newcommand{\ccuavii}{\mbox{\usebox{\cuavii} \hskip 15pt  }
} 

\newsavebox{\cuaviii}
\savebox{\cuaviii}(10,3){
\put(2,5) {\circle{10}}
\put(-1,0.8){\line(0,1){8.2}}
\put(2,0){\line(1,3){2.99} }
\put(5,1){\line(-1,3){3} }
\put(-3,5){\line(3,0){10} }}
\newcommand{\ccuaviii}{\mbox{\usebox{\cuaviii} \hskip 15pt  }
}

\newsavebox{\cuaix}
\savebox{\cuaix}(10,3){
\put(2,5) {\circle{10}}
\put(0,0.3){\line(0,1){9.2}}
\put(4,0.3){\line(0,1){9.2} }
\put(-2.7,3){\line(3,0){9.2}}
\put(-2.7,7){\line(3,0){9.2}} }
\newcommand{\ccuaix}{\mbox{\usebox{\cuaix} \hskip 15pt  }
} 

\newsavebox{\cuax}
\savebox{\cuax}(10,3){
\put(2,5) {\circle{10}}
\put(-0.3,0.5){\line(1,2){4.5}}
\put(4.3,0.6){\line(-1,2){4.45} }
\put(-2.7,3){\line(3,0){9.2}}
\put(-2.7,7){\line(3,0){9.2}} }
\newcommand{\ccuax}{\mbox{\usebox{\cuax} \hskip 15pt  }
} 

\newsavebox{\cuaxi}
\savebox{\cuaxi}(10,3){
\put(2,5) {\circle{10}}
\put(-3,5){\line(3,0){10}}
\put(2,0){\line(0,5){10} }
\put(-1.6,1.4){\line(1,1){7.1} }
\put(5.5,1.5){\line(-1,1){7.05} }}
\newcommand{\ccuaxi}{\mbox{\usebox{\cuaxi} \hskip 15pt  } } 

\newsavebox{\extra}
\savebox{\extra}(10,3){
\put(2,5) {\circle{10}}
\put(0,0.3){\line(0,1){9.2}}
\put(4,0.4){\line(0,1){9.1} } }
\newcommand{\cextra}{\mbox{\usebox{\extra} \hskip 15pt  }
}

\begin{document}
\begin{titlepage}

\begin{flushright}
USC-FT-7/99 \\
hep-th/9905057 
\end{flushright}
\vglue 1cm

\begin{center}

{\large \bf  CHERN-SIMONS GAUGE THEORY:} 
\vskip5pt
{\large \bf TEN YEARS AFTER\footnotemark}

\footnotetext{Invited lecture delivered at the workshop ``Trends in
Theoretical Physics II", held at Buenos Aires, Argentina, November
29 -- December 5, 1998.}

\vskip1cm

{\large J. M. F. Labastida}

\vspace{0.5cm}

{\em Departamento de F\'\i sica de Part\'\i culas \\
Universidade de Santiago de Compostela \\
E-15706 Santiago de Compostela, Spain\\
{\rm e-mail: labasti@fpaxp1.usc.es}}

\end{center}

\vskip1.5cm

\begin{abstract}

A brief review on the progress made in the study of Chern-Simons
gauge theory since its relation to knot theory was discovered ten years ago
is presented. Emphasis is made on the analysis of the perturbative study of
the theory and its connection to the theory of Vassiliev invariants. It is
described how the study of the quantum field theory for three different gauge
fixings leads to three different representations for Vassiliev invariants.
Two of these gauge fixings lead to well known representations: the covariant
Landau gauge corresponds to the configuration space integrals while the
non-covariant light-cone gauge to the Kontsevich integral. The progress made
in the analysis of the third gauge fixing, the non-covariant temporal gauge,
is described in detail. In this case one obtains combinatorial expressions,
instead of integral ones, for Vassiliev invariants. The approach based on
this last gauge fixing seems very promising to obtain a full combinatorial
formula. We collect the combinatorial expressions for all the Vassiliev
invariants up to order four which have been obtained in this approach. 

\end{abstract}

\vbox{\vskip2cm}

\vskip2cm
\vfill

\end{titlepage}

\def\theequation{\thesection.\arabic{equation}}

\section{Introduction}
\setcounter{equation}{0}

The connection between Chern-Simons gauge theory and the theory of
knot and link invariants was established by Edward Witten ten
years ago \cite{csgt}. Since then the theory has been studied from a variety
of points of view. Many of the standard methods in field theory have
been applied, generating results which became 
important in the development of knot theory. 
The interplay between quantum field theory and knot theory has been very
rich in both directions. Though the results have been more spectacular in the
knot theory direction, one must not forget that the
developments in Chern-Simons gauge theory have constituted a constant test of
our knowledge in quantum field theory. In fact, it has been found that
not always the quantum field theory methods have been able to provide the
right answer. As it will be described in detail, the work of the last few
years reveals that there are some issues which are not yet understood when
dealing with non-covariant gauges. 

The term Chern-Simons theory appears in different contexts of
quantum field theory. In this conference these words have been
heard at least in half of the talks. It is therefore convenient to
specify which type of Chern-Simons gauge theory I will be dealing
with in this paper. I will refer by the term Chern-Simons
gauge theory to a quantum field theory in three-dimensions whose
action is the integral on a smooth compact boundaryless three-manifold of a
Chern-Simons form associated to  a semi-simple non-abelian gauge group. This
theory was originally considered by several authors \cite{primeros}, but only
after the work by Witten its connection with the theory of knot and link
invariants was discovered. This occurred in the summer of 1988. Some of the
other theories also named by the term Chern-Simons have been recently reviewed
in \cite{dunne}.

The presentation contained in this paper will not follow a
chronological order. The development of the theory of knot and link
invariants from a mathematical point of view in the last fifteen
years have been very impressive and at some stages it has developed
parallel to  Chern-Simons gauge theory. I will
try to make a correspondence from each side for each of the topics treated.
Though I have witnessed the development of Chern-Simons gauge theory in
detail during these  ten years, I might have missed some of the
corresponding achievements from the mathematical side. I apologize in
advance if some omission in this respect takes place. The order in the
presentation is chosen in such a way that the progress made in these ten
years can be understood starting from the basics of both, knot and
Chern-Simons gauge theory. At each stage the results obtained in the context
of Chern-Simons gauge theory are interpreted in the context of knot theory
from a mathematical point of view.

The study of Chern-Simons gauge theory is an unusual one because
it was first analyzed from a non-perturbative point of view. The
original paper by Witten presents a series of non-perturbative
methods which led him to show that the vacuum expectation values
(vevs) of the relevant operators of the theory are polynomial
invariants like the Jones  polynomial \cite{jones} and its generalization.
Other non-perturbative analysis were made one year later and soon some of
the first perturbative studies started to appear. However, only
some years later, after the advent of Vassiliev invariants \cite{vass,birrev},
the importance of the study of the perturbative series expansion was
recognized. From a field theory point of view this lack of interest
was understandable. Usually, field theorist can grasp only some of
the perturbative aspects of their theories. Since in Chern-Simons
gauge theory we had a good handle on its exact solution, why
should one care about its perturbative series?  It turned out that
the coefficients of the perturbative series are important
invariants. Their study applying perturbation theory led
to interesting expressions for them. The invariance of these
coefficients was clear from the beginning. What was not obvious is
that they were invariants with a very special feature, they were
Vassiliev invariants or invariants of finite type. Though this was
known since the work by Bar-Natan \cite{barnatan} and Birman and Lin
\cite{bilin} in 1993, its proof in a quantum field theory context had to wait
until 1997 \cite{singular}.

Once the importance of the perturbative series expansion was
realized several works addressed its study. Again the richness
inherent to quantum field theory become a powerful tool and the
form of this series expansion was studied for different gauge-fixings. The
pioneer perturbative calculations in the covariant Landau gauge
\cite{gmm,natan}
were later extended and analyzed from a general point of view
\cite{alla,alts}. All these works constituted part of the inspiration for the
formulation by Bott and Taubes of their configuration space integral
\cite{bt}. Their integral corresponds precisely to the perturbative series
expansion of the vev of a Wilson loop in Chern-Simons gauge theory in the
Landau gauge. Before the work by Bott and Taubes, Kontsevich
presented a different integral \cite{kont} for Vassiliev invariants. This
integral turned out to correspond to the perturbative series 
expansion of the vev of a Wilson loop in the non-covariant light-cone
gauge \cite{cata,lcone,kaucone}. The interplay between physics and mathematics
was very fruitful in these developments. This is rather clear in the case
of the covariant Landau gauge. In the case of the light-cone
gauge, the Chern-Simons counterpart took place much later than the
formulation of the Kontsevich integral. However, as stated in the
paper by Kontsevich, some of his insight to write down his
integral originated from Chern-Simons gauge theory. A very recent
work \cite{poir}  shows from a mathematical point of view that both, 
the Kontsevich integral and the Bott and Taubes configuration
space integral lead to the same invariants. From a field theory
point of view this is just a consequence of the gauge invariance of
the theory.

Gauge invariance is a powerful tool: it allows to study the
theory for different gauge fixings. In the last few years a new
gauge fixing has been considered: the non-covariant temporal gauge
\cite{cata,temporal}.
This gauge has the important feature that the integrals which are
present in the expressions for the coefficients of the perturbative
series expansion can be carried out, leading to combinatorial
expressions \cite{temporal}. This has been shown to be the case up to order
four and it seems likely that the approach can be generalized. In this
analysis a crucial role is played by the factorization theorem for
Chern-Simons gauge theory proved in
\cite{factor}. The resulting expressions are better presented when
written in terms of  Gauss diagrams for knots \cite{faroknots}. Some recent
results from the mathematical side seem to indicate the  existence of a
combinatorial formula of this type \cite{virgpo}. At present Chern-Simons gauge
theory is the only approach which have provided combinatorial
expressions for all the Vassiliev invariants up to order four.
Work is in progress from both sides to obtain a general
combinatorial expression. Hopefully, a coherent interplay between
them will provide the widely searched general combinatorial
expression for Vassiliev invariants.

The paper is organized  using the following table as a guide. I have
listed on the left hand side the mathematical counterparts of the
topics of Chern-Simons gauge theory listed on the right hand side.

\vskip5pt
\begin{table}[hp]
\begin{center}
\begin{tabular}{||c||c||}
\hline {\bf Knot Theory} & {\bf Chern-Simons Gauge Theory} \\
\hline\hline Knots and links & Wilson loops \\
\hline Knot and link polynomial invariants &Vevs of products of Wilson
loops \\
\hline Singular knots & Operators for singular knots \\
\hline Invariants for singular knots & Vevs of the new operators  \\
\hline Finite type or Vassiliev invariants & Coeffs. of the
perturbative series \\
\hline Chord diagram & First coeff. of the perturbative series \\
\hline \{1T,4T\} and \{1T,AS,IHX,STU\} & Lie-algebra structure of group
factors \\
\hline Configuration space integral & Landau gauge \\
\hline Kontsevich integral & Light-cone gauge \\
\hline ?? & Temporal gauge \\
\hline
\end{tabular}
\end{center}
\end{table}

\vskip-9pt
\noindent The sections of the paper will deal with the development of
these topics from the point of view of Chern-Simons gauge theory, indicating
the basic details from its knot theory counterpart. Notice that the entry in
the knot-theory column corresponding to the temporal gauge has not been filled
in yet.

The paper is organized as follows. In sect. 2, Chern-Simons
gauge theory is introduced, as well as the basics of the theory of knots
and links. A brief summary of some of the results obtained in the context of
non-perturbative Chern-Simons gauge theory is also included in this section. In
sect. 3, singular knots are introduced and their corresponding operators are
defined. These operators are studied and their connection to chord diagrams is
discussed. In sect. 4, Chern-Simons gauge theory is studied from a
perturbative point of view. Using the covariant Landau gauge the coefficients
of the perturbative series expansion are analyzed, obtaining configuration
space integrals. In sect. 5, the perturbative series expansion is reobtained
in the non-covariant light-cone gauge. The resulting series is the same as
the Kontsevich integral for Vassiliev invariants. In sect. 6, the perturbative
series expansion is analyzed in a different non-covariant gauge, the temporal
gauge. The general procedure to obtain combinatorial expressions for Vassiliev
invariants is presented and their explicit form is given for all the
primitive invariants up to order four. Finally, in sect. 7,
 some
concluding remarks are presented. Two tables list all the primitive Vassiliev
invariants up to order four for all prime knots up to nine crossings.

\vfill
\newpage

\section{Knots, links and Wilson loops}
\setcounter{equation}{0}

Let us begin recalling the basic elements of Chern-Simons gauge theory. This
theory is a quantum field theory whose action is based on the Chern-Simons
form associated to a non-abelian gauge group. 
The fundamental data in Chern-Simons gauge theory are the following: a
smooth three-manifold $M$ which  will be taken to be compact, a gauge
group $G$, which will be taken semi-simple and compact, and an integer
parameter $k$.  The action of the theory is the
integral of the Chern-Simons form associated to a gauge connection $A$
corresponding to a gauge group $G$:
\begin{equation}
S_{\CS} (A) ={k\over 4\pi} \int_M \tr (A\wedge d A + \frac{2}{3} A\wedge
A\wedge A).
\label{valery}
\end{equation}
In this expression the trace is taken in the fundamental representation of the
gauge group. The action possesses the following behavior under gauge
transformations,
\begin{eqnarray}
A &\rightarrow & A + g^{-1} d g, \\
S_{\CS} (A) &\rightarrow & S_{\CS} (A) - 4 \pi  y k \omega(g),
\label{jane}
\end{eqnarray}
where $g$ is a map $g: M \rightarrow G$, $\omega(g)$ is its winding number,
\begin{equation}
\omega(g) = {1\over 48 \pi ^2 y}\int_M
\tr(g^{-1} d g\wedge g^{-1} d g\wedge g^{-1} d g),
\label{clara}
\end{equation}
and $y$ the Dynkin index of the fundamental representation of $G$.
Since the winding number (\ref{clara}) is an integer and $y$ is a half-integer,
it follows  that, for integer $k$, the exponential,
\begin{equation}
\exp(i S_{\CS}(A) ),
\label{pregreta}
\end{equation}
which is the quantity that enters in the computation of vevs, is invariant
under gauge transformations.

A theory characterized by an action like (\ref{valery}), which is independent
of the metric on the three-manifold $M$, is a topological quantum field
theory. In this theory, appropriate observables lead to vevs which
correspond to topological invariants. Candidates to be observables of this
type have to satisfy two properties. On the one hand they must be metric
independent, on the other hand they must be gauge invariant. Wilson loops
verify these two properties and they are therefore the paramount observables
to be considered in Chern-Simons gauge theory.

Wilson loop operators correspond to the holonomy of the gauge connection $A$
along a loop. Given a representation $R$ of the gauge group $G$ and a 1-cycle
$\gamma$ on $M$, it is defined as:
\begin{equation}
W_\gamma^R (A) =\tr_R \big( {\hbox{\rm Hol}}_\gamma (A) \big) =
\tr_R {\hbox{\rm P}} \exp \int_\gamma A.
\label{rosalia}
\end{equation}
Products of these operators are the natural candidates to obtain
topological invariants after computing their vev. These
vevs are formally written as: 
\begin{equation}
\langle W_{\gamma_1}^{R_1} W_{\gamma_2}^{R_2} \cdots
W_{\gamma_n}^{R_n} \rangle 
        \nonumber \\
= \int [DA] W_{\gamma_1}^{R_1}(A) W_{\gamma_2}^{R_2}(A) \cdots
W_{\gamma_n}^{R_n}(A)   \ex^{i S_{\CSs} (A) },
\label{encarna}
\end{equation}
where $\gamma_1$, $\gamma_2$,$\dots$,$\gamma_n$ are 1-cycles on $M$ and
$R_1$,
$R_2$ and $R_n$ are representations of $G$. In (\ref{encarna}), the quantity
$[DA]$ denotes the functional integral measure and it is assumed that an
integration over connections modulo gauge transformations is carried out. As
usual in quantum field theory this integration is not well defined. Field
theorists have elaborated a variety of methods to go around this problem and
provide some meaning to the right hand side of (\ref{encarna}). These methods
fall into two categories, perturbative and non-perturbative ones, and their
degree of success mostly depends on the quantum field theory under
consideration. Fortunately, in Chern-Simons gauge theory these methods have
been very fruitful.  Indeed, in the pioneer work by Witten in 1988 he showed,
using non-perturbative methods, that when one considers non-intersecting cycles
$\gamma_1$, $\gamma_2$,$\dots$,$\gamma_n$ without self-intersections, the
vevs (\ref{encarna}) lead to the polynomial invariants
discovered a few years before starting with the work by V. F. Jones
\cite{jones}. But before making a precise statement on what was achieved by
Witten in
\cite{csgt} let us go through our first mathematical detour and collect some
basic facts on knot theory.

\begin{figure}
\centerline{\epsffile{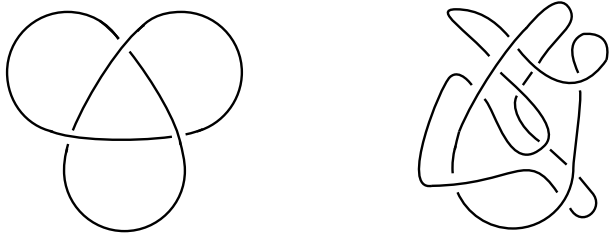}}
\caption{Two regular projections of the trefoil knot.}
\label{trefoil}
\end{figure}

Knot theory studies embeddings $\gamma:$ $S^1 \rightarrow M$. Two of these
embeddings are considered equivalent if the image of one of them can be
deformed into the image of the other by an homeomorphism on $M$. The main goal
of the theory is to classify the resulting equivalence classes. Each of these
classes is a knot. Most of the study in knot theory has been carried out for
the simple case
$M=S^3$. This is the situation which has been  widely studied from a
Chern-Simons theory point of view. Chern-Simons gauge theory, however, being a
formulation intrinsically three-dimensional, provides a framework to study the
case of more general three-manifolds $M$. In this respect, Chern-Simons gauge
theory seems more promising than other approaches whose formulation possesses a
two-dimensional flavor.

A powerful approach to classify knots is based on the construction of knot
invariants. These are quantities which can be computed taking a
representative of a class and are invariant within the class, \ie, are
invariant under continuous deformations of the representative chosen. At
present, it is not known if there exist enough knot invariants to classify
knots. Vassiliev invariants are the most promising candidates but is already
known that if they do classify, infinitely many of them are needed. 

\begin{figure}
\centerline{\epsffile{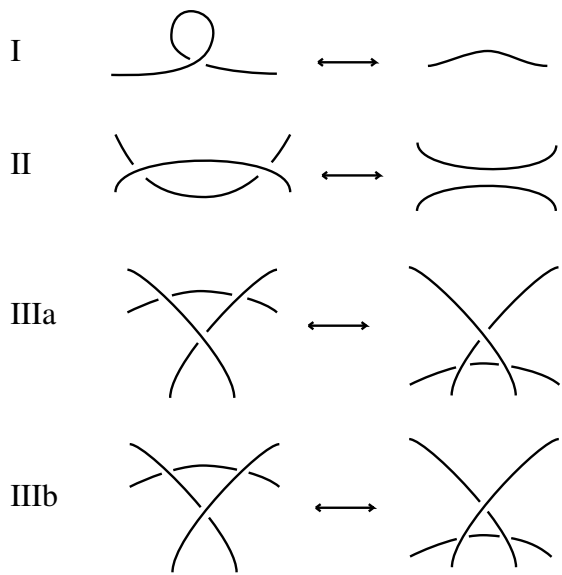}}
\caption{Reidemeister moves.}
\label{reide}
\end{figure}

The problem of the classification of knots in $S^3$ can be reformulated in a
two-dimensional framework using regular knot projections. Given a
representative of a knot in $S^3$, deform it continuously in such a way that
the projection on a plane has simple crossings. Draw the projection on the
plane and at each crossing use the convention that the line that goes under
the crossing is erased in a neighborhood of the crossing. The resulting
diagram is a set of segments on the plane, containing the relevant
information at the crossings. Two diagrams corresponding to two regular knot
projections of the trefoil knot are shown in fig. \ref{trefoil}. A given knot
might have many regular projections. The first question to ask is if one can
define an equivalence relation among regular projections whose equivalence
classes coincide with the equivalence classes of knots. Reidemeister answered
this question in the affirmative many years ago. He proved that the problem of
classifying knots was equivalent to the problem of classifying knot
projections modulo a series of relations among them. These relations, known as
Reidemeister moves, are the ones shown in fig. \ref{reide}. Invariance of a
quantity under the three Reidemeister moves is called invariance under ambient
isotopy. If a quantity is invariant under all but the first is said to possess
invariance under regular isotopy. It is instructive to find out the way that
the Reidemeister moves can be applied to the two diagrams shown in fig.
\ref{trefoil} to show that, indeed, they are equivalent.

The formalism described for knots  generalizes to the case of links.
For a link of $n$ components  one considers  $n$ embeddings,
$\gamma_i:$ $S^1\rightarrow M$, $i=1,\dots,n$, with no intersections among
them. Again, the main problem that link theory faces is the problem of their
classification modulo homeomorphisms on $M$. In this case one can also define
regular projections and reformulate the problem in terms of their
classification modulo the Reidemeister moves shown in fig. \ref{reide}.
Notice that $n=1$ is just the case which corresponds to knots.

The study of knot and link invariants experimented important progress in
the eighties. In 1984 V. F. Jones \cite{jones} discovered a new invariant
which strongly influenced the field.  He formulated the celebrated Jones
polynomial, an invariant which was able to distinguish many more knots and
links that previous knot invariants. To have a flavor of the type of
quantities one is dealing with, let us describe it in some detail. 

\begin{figure}
\centerline{\epsffile{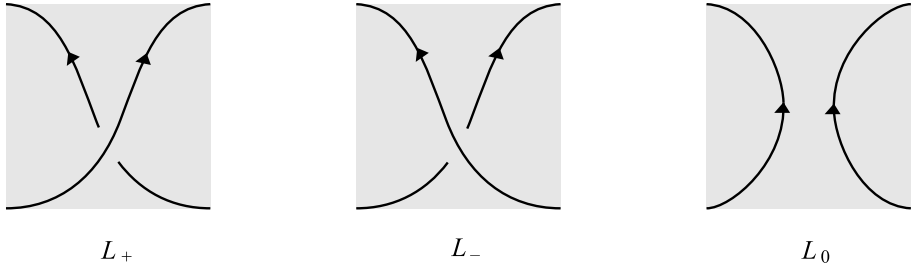}}
\caption{Links which enter in the skein rule of the Jones polynomial.}
\label{skein}
\end{figure}

The Jones polynomial can be defined very simply in terms of skein relations.
These are a set of rules that can be applied to the diagram of a regular knot
projection to construct the polynomial invariant. They establish a relation
between the invariants associated to three links which only differ in a region
as shown in fig. \ref{skein}. Notice that arrows have been provided to each of
the segments entering fig. \ref{skein}. Indeed, the Jones polynomial as
well as, in general,  the rest of polynomial invariants which will be
discussed below are defined for oriented links. Thus, an arrow must be
introduced for each of the components of a link. Though polynomial knot
invariants are invariant under a reversal of its orientation, link polynomial
invariants are not.

If one denotes by $V_{L}(t)$ the Jones polynomial corresponding to a link
$L$, being $t$ the argument of the polynomial, it must satisfy the skein
relation:
\begin{equation}
{1\over t}V_{L_+} - t V_{L_-} = (\sqrt{t} - {1\over \sqrt{t}})V_{L_0},
\label{jonespoli}
\end{equation}
where $L_+$, $L_-$ and $L_0$ are the links pictured in fig. \ref{skein}.
This relation plus a choice of normalization for the unknot ($U$) are enough
to compute the Jones polynomial for any link. The standard choice for the
unknot is:
\begin{equation}
V_{U}=1,
\label{nonudo}
\end{equation}
though it is not the most natural one from the point of view of Chern-Simons
gauge theory. For the trefoil knot shown in fig. \ref{trefoil} this polynomial
is: $V_T=t+t^3-t^4$. Notice that actually $V_L(t)$, in general, is not a
polynomial. First, it might contain negative powers of $t$. Indeed, for the
mirror image of the trefoil knot shown in fig. \ref{trefoil}, $\tilde T$, one
easily finds that $V_{\tilde T}=t^{-1}+t^{-3}-t^{-4}$. Actually this is just
an example of a general feature of the Jones polynomial that under a reversal
of the orientation of the ambient space $S^3$ it behaves as $V_{\tilde
L}(t)=V_L(t^{-1})$, a property which makes it stronger than the Alexander
polynomial which was not able two discriminate among links $L$ and $\tilde L$ 
which are mirror images of each other. Second, $V_L(t)$ might contain a
factor of the form
$\sqrt{t}$. This is always the case if $L$ has an even number of components.
For example, for the simplest two-component link, the Hopf link, it takes the
form: $V_H=\sqrt{t}(1+t^2)$.

After Jones work in 1984, many other  polynomial invariants where discovered.
Two of the most celebrated ones are the HOMFLY \cite{homfly} and the Kauffman
\cite{kauffman} polynomial invariants. The first one possesses a skein rule
with three entries similar to the one in fig. \ref{skein} and can be computed
in the same way. The novelty is that it is a polynomial in two variables. The
second one is also a polynomial in two variables but its corresponding skein
rule is not as simple as in the Jones or HOMFLY polynomials. In general, the
generalizations involve  skein rules containing more than three terms
and the computation of invariants becomes more complicated. Often other
methods have to be used. Before entering in the discussion of the general
framework which accounts for all these developments let us turn back to
Chern-Simons gauge theory.

The pioneer work by Witten in 1988 showed that the vevs of products of Wilson
loops (\ref{encarna}) correspond to the Jones polynomial when one considers
$SU(2)$ as gauge group and all the Wilson loops entering in the vev are taken
in the fundamental representation $F$. For example, if one considers a knot
$K$, Witten showed that,
\begin{equation}
V_K(t)= \langle W_{K}^{F} \rangle,
\label{casouno}
\end{equation}
provided that one performs the identification:
\begin{equation}
t= \exp({2\pi i\over k+h}),
\label{iden}
\end{equation}
where $h=2$ is the dual Coxeter number of the gauge group $SU(2)$. Witten
also showed that if instead of $SU(2)$ one considers $SU(N)$ and the Wilson
loop carries the fundamental representation, the resulting invariant is the
HOMFLY polynomial. The second variable of this polynomial originates in this
context from the $N$ dependence. But these cases are just a sample of the
general framework which Chern-Simons gauge theory offers. Taking other groups
and other representations one possesses an enormous set of knot and link 
invariants. For some other special cases the resulting invariants correspond
to specific polynomial invariants. For example, in one considers $SO(N)$ as
gauge group and  Wilson loops carrying the fundamental representation one is
led to the Kauffman polynomial. If instead one considers $SU(2)$ as gauge
group and the Wilson loops carry a representation of spin $j/2$ one
rediscovers the Akutsu-Wadati \cite{aku} polynomials. Notice that the
formalism allows to consider different representations for each of the
components of a link, leading to the so-called colored polynomial invariants.

Chern-Simons gauge theory constitutes a framework where an enormous
variety of knot and link invariants can be considered. The theory can be
studied for arbitrary groups and arbitrary representations. But on top of
this it also provides the basis to study these invariants in more general
three-manifolds. In addition, Chern-Simons gauge theory also allows to
consider a more general set of observables called graphs \cite{wittengraphs}
that certainly constitutes an important generalization which has not been much
exploited from the field theory point of view.

The framework established from Chern-Simons gauge theory has been also
formulated from a mathematical point of view. In general, the invariants
inherent to Chern-Simons gauge theory have been reobtained and properly
defined using contexts different than quantum field theory. This has been a
very fruitful arena in the field of algebraic topology in the last ten
years. There exist now a quantum group approach to polynomial invariants
\cite{qga}, and the general theory, including knots, links and graphs has been
formulated at a categorical level \cite{cate}.

A particular invariant of three-manifolds which deserves special attention is
the partition function of Chern-Simons gauge theory. This quantity is hard to
obtain from a field theory point of view. However, it has been properly
defined from a mathematical point of view using triangulations of the
three-manifold. The resulting invariant is known as the
Witten-Reshetekhin-Turaev invariant \cite{wrt}, and it corresponds to the
mentioned partition function. The partition function
has been studied for some three-manifolds in \cite{freed}. The
Witten-Reshetekhin-Turaev invariant can be obtained from Chern-Simons gauge
theory using lattice gauge theory methods and placing the quantum field
theory on a triangulated three-manifold
\cite{gambiniuno}.  Recently, new
developments based in Chern-Simons gauge theory has led to new formulae for
the partition function \cite{kaultv}. The approach seems very promising and it
might lead to simpler computational methods for these invariants.

\begin{figure}
\centerline{\epsffile{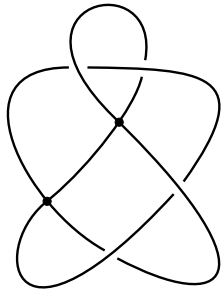}}
\caption{Regular projection of a singular knot with two double
points.}
\label{singu}
\end{figure}

Many non-perturbative studies of Chern-Simons gauge theory were performed in
the years following Witten's seminal work. The quantization of the theory was
studied from the point of view of the operator formalism
\cite{bos,opf,eli} and from more geometrical methods \cite{pietra}.
Also, its connection to two-dimensional conformal field theory was further
elucidated \cite{king} .A powerful method for the general computation of knot
and link invariants was constructed by Kaul and collaborators
\cite{kaul}. Methods to compute graph invariants were also built
\cite{martin}. All these works provided good setups for calculation purposes
that in some situations were able to provide the answer to some open questions
in knot theory. For example, general expressions for torus knots and torus
links were obtained for a variety of situations using the operator formalism
\cite{torus}. The problem of finding a polynomial invariant which
discriminates between the two chiralities of the knots
$9_{42}$ and $10_{71}$ was solved \cite{chir} using the methods developed by
Kaul and collaborators. This approach was also used to show that polynomial
invariants do not distinguish isotopically inequivalent mutant knots and links
\cite{mutant}. The connection between Chern-Simons gauge theory and rational
conformal field theory was used to build knot and link invariants from any
conformal field theory \cite{ratious,ratio}. Chern-Simons gauge theory has had
also important applications in the loop-representation approach to canonical
quantum gravity
\cite{rovsmo,bgp}. Recently, graphs have also become very important in this
context and it turns out that their associated Vassiliev invariants are related
to physical states in the framework of canonical quantum gravity \cite{ggp}.

\vfill
\newpage

\section{Singular knots and their operators}
\setcounter{equation}{0}

In this section  the mathematical point of view will be discussed first. This
will motivate the need to consider vevs of new operators
associated to loops with self-intersections. We will describe how
Chern-Simons gauge theory leads, using quantum field theory techniques, to
the same results as its mathematical counterpart.

In 1990, V. A.  Vassiliev \cite{vass} introduced a new point of view to study
the problem of the classification of knots. Based on Arnold's work on
singularity theory he studied the space of all smooth maps of $S^1$ into
$S^3$. This includes maps with various types of singularities which divide the
space into chambers, each corresponding to a knot type. Using methods of
spectral sequences to obtain combinatorial conditions, Vassiliev constructed
families of new invariants to characterize these chambers.

\begin{figure}
\centerline{\epsffile{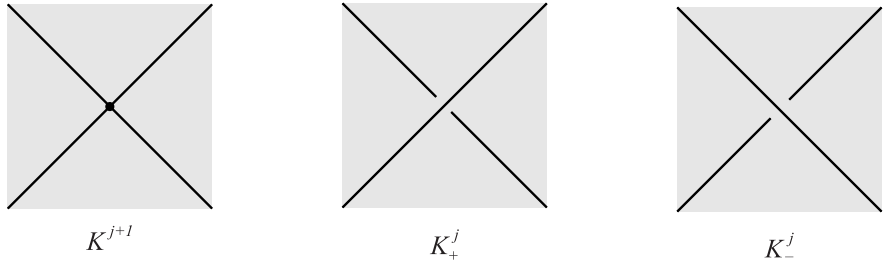}}
\caption{Regular projections which enter in the Vassiliev resolution of a
double point.}
\label{vassre}
\end{figure}

Vassiliev approach was later reformulated by Birman and Lin \cite{bilin} from
an axiomatic point of view. As in the original formulation by
Vassiliev \cite{vass}, the starting point is based on the consideration 
of singular knots with $j$ double points. A representative of a singular knot
with $j$ double points consists of the image of a map from $S^1$ into $S^3$
with
$j$ simple self-intersections. Under homeomorphisms on $S^3$ these images form
a class which constitute the singular knot itself. These singular knots can
also be  regularly projected. Fig. \ref{singu} shows one of these knots with
two double points. The key ingredient in the construction by Birman and Lin is
the observation that any knot invariant extends to generic singular knots by
the Vassiliev resolution:
\begin{equation}
\nu({K^{j+1}}) =\nu({K^j_+}) - \nu({K^j_-}),
\label{vassreso}
\end{equation}
where $K^{j+1}$ is a singular knot with $j+1$ double points which differs
from the knots $K^j_+$ and  $K^j_-$ only in the region shown in fig.
\ref{vassre}.
Using this extension Birman and Lin characterized the invariants of finite
type or Vassiliev invariants introducing the following definition:

 A Vassiliev or finite type invariant of order $m$ is a knot invariant
which is zero on the unknot and that, after extending it to singular knots,
it is zero on singular knots $K^j$ with $j>m$ singular points, and different
from zero on some $K^m$.

Vassiliev invariants form  a vector space ${\cal V}$. Linear
combinations of Vassiliev invariants are also Vassiliev invariants. Actually,
to a Vassiliev invariant of order
$m$ one can always add Vassiliev invariants of lower order obtaining
additional Vassiliev invariants of order
$m$. It is convenient to consider a filtration in the vector space of
Vassiliev invariants using the order as grading. If one denotes by ${\cal
V}_m$ the space of Vassiliev invariants of order $m$ one has the filtration,
\begin{equation}
{\cal V}_0 \subseteq {\cal V}_1 \subseteq \cdots \subseteq
{\cal V}_m \subseteq \cdots \subset {\cal V},
\label{contenido}
\end{equation}
which leads to the graded vector space:
\begin{equation}
{\rm gr}({\cal V})=
{\cal V}_0 \oplus {\cal V}_1 / {\cal V}_0 \oplus {\cal V}_2 / {\cal V}_1
\oplus \cdots \oplus {\cal V}_{m+1} / {\cal V}_m \oplus \cdots
\label{graded}
\end{equation}

\begin{figure}
\centerline{\epsffile{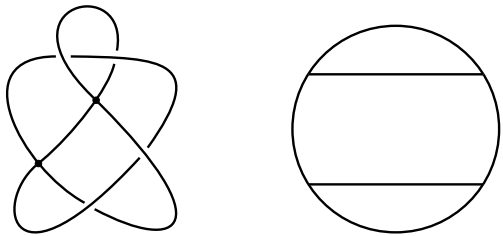}}
\caption{A singular knot and its corresponding chord diagram.}
\label{cdiag}
\end{figure}

Besides introducing an axiomatic approach to Vassiliev invariants, Birman and
Lin proved an important theorem in 1993 \cite{bilin}. Let us consider any
polynomial invariant $P_K(t)$ for a knot $K$. This polynomial could be any of
the ones obtained from Chern-Simons gauge theory considering the vev of the
corresponding Wilson loop for some group and some representation. Consider
now the power series expansion:
\begin{equation}
Q_K(x) = P_K(\ex^x) = \sum_{m=0}^\infty \nu_m (K) x^m.
\label{elteo}
\end{equation}
Birman and Lin proved that if one extends the quantities $\nu_m(K)$ to
Vassiliev invariants for singular knots using Vassiliev resolution
(\ref{vassreso}), then $\nu_m(K)$ are Vassiliev invariants of order $m$. An
immediate consequence of this theorem is that the coefficients of the
perturbative expansion associated to the vev of a Wilson loop in Chern-Simons
gauge theory are Vassiliev invariants. This property of the coefficients of
the perturbative series expansion has been proved using standard quantum
field theory methods
\cite{singular}. Before describing how this has been achieved let us
make some remarks on Vassiliev invariants and introduce one of them which is
particularly important.

Once Vassiliev or finite type invariants have been introduced, there are two
basic questions to ask. The first one is whether or not all the known
numerical knot invariants are of finite type. The answer to this question is
no. There are classical numerical knot invariants like the unknotting number,
the genus or the crossing number which are not Vassiliev invariants. The second
question is whether or not, in the case that Vassiliev invariants classify
knots, one needs an infinite sequence of Vassiliev invariants to separate
knots. This question has been answer in the affirmative.

From a singular knot with $m$ double points one can construct a
particular object which determines Vassiliev invariants of order
$m$: its chord diagram \cite{barnatan}. Given a singular knot $K^m$, its chord
diagram, $CD(K^m)$, is built in the following way. Take a base point and draw
the preimages of the map associated to a given representative of $K^m$ on a
circle. Then join by straight lines the pairs of preimages which correspond to
each singular point. In fig. \ref{cdiag}  the chord diagram
corresponding to the singular knot of fig. \ref{singu} has been pictured.
It is rather simple to observe that if $\nu(K^m)$ is a Vassiliev invariant of
order $m$ then it is completely determined by $CD(K^m)$. Indeed, if one
considers two singular knots with $m$ double points, $K^m_1$ and $K^m_2$,
which differ in one crossing change then $CD(K^m_1)=CD(K^m_2)$. On the other
hand, by Vassiliev resolution $\nu(K^m_1)-\nu(K^m_2)=\nu(K^{m+1})$. But
$\nu(K^{m+1})=0$ and therefore all singular knots with $m$ double points
leading to the same chord diagram have the same Vassiliev invariant of order
$m$.

\begin{figure}
\centerline{\epsffile{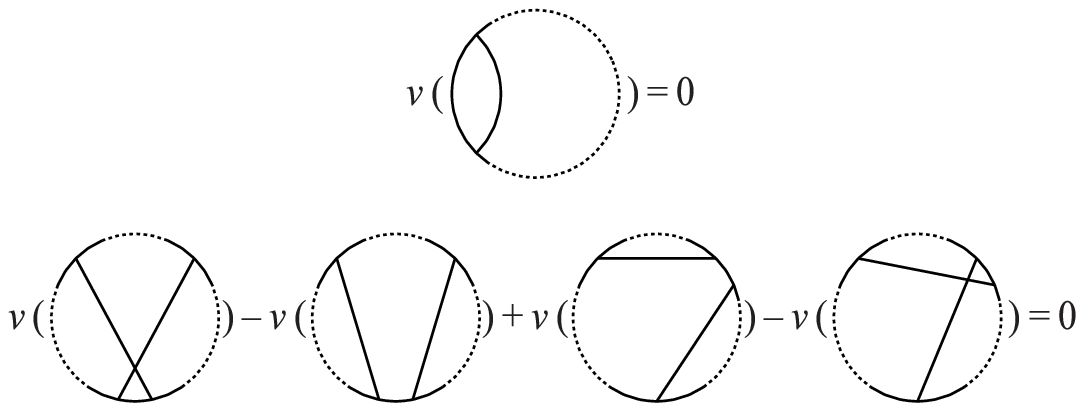}}
\caption{1T and 4T relations among chord diagrams.}
\label{fourt}
\end{figure}

Chord diagrams play an important role in the theory of Vassiliev invariants.
Since Vassiliev invariants of order $m$ for singular knots with $m$ double
points are codified by chord diagrams one could ask if there are as many
independent invariants of this kind as chord diagrams. The answer to this
question is no. Chord diagrams are associated to knot diagrams
and these diagrams must be considered modulo the equivalence relation
dictated by the Reidemeister moves in fig. \ref{reide}. These relations indeed
impose some relations among chord diagrams, the so-called 1T and 4T relations
\cite{barnatan}. They have been depicted in fig. \ref{fourt}. 

Of particular importance is the space of chord diagrams modulo the 1T and 4T
relations. This space is a graded space, graded by the number of chords:
\begin{eqnarray}
{\cal A} &=& \Big\{ {\rm chord\,\,\, diagrams} \Big\} \,\, \Big/
\,\, {\rm (1T,\,\, 4T)}
\nonumber \\
&=& {\cal A}_0 \oplus {\cal A}_1 \oplus \cdots \oplus {\cal A}_m \oplus
\cdots
\label{chordsp} 
\end{eqnarray}
Notice that ${\cal A}_m$ labels all the Vassiliev invariants of order $m$ for
all knots with $m$ double points. The dimensions of the vector spaces
${\cal A}_m $ are not known in general. The space ${\cal A}$ possesses an
algebraic structure that reduces the study of the spaces
${\cal A}_m$ to its connected part $\hat{\cal A}_m$. If $\hat d_m$ denotes
the dimension of the connected part of ${\cal A}_m$, it is known that they
take the following values up to order 12 \cite{jj}:

\begin{table}[htbp]
\centering
\begin{tabular}{|c|c|c|c|c|c|c|c|c|c|c|c|c|}  \hline
 $m$ & 1 & 2 & 3 & 4 & 5 & 6 & 7 & 8 & 9 & 10 & 11 & 12 \\  \hline
 $\hat d_m$ & 0 & 1 & 1 & 2 & 3 & 5 & 8 & 12 & 18 & 27 & 39 & 55 \\ 
\hline
\end{tabular}
\label{latablados}
\end{table}

\noindent The general expression for the dimensions of the spaces  of chord
 diagrams $\hat{\cal A}_m$ is an open problem which has challenged many
people. As it will be discussed below, these dimensions correspond in fact to
the dimensions of the spaces of primitive Vassiliev invariants.

\begin{figure}
\centerline{\epsffile{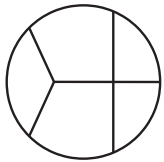}}
\caption{Diagram with one trivalent vertex.}
\label{tresv}
\end{figure}

\begin{figure}
\centerline{\epsffile{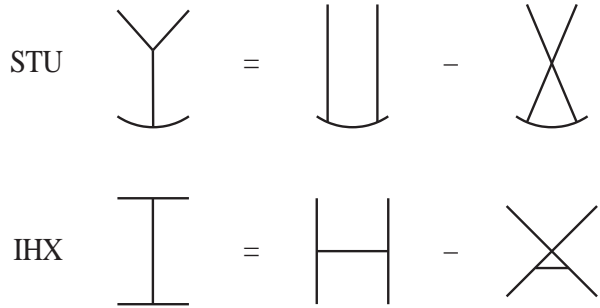}}
\caption{STU and IHX relations among trivalent diagrams.}
\label{stu}
\end{figure}

The vector space of chord diagrams, ${\cal A}$, can be characterized in an
equivalent way using trivalent diagrams an introducing a series of new
relations. This characterization is very important because, as it will be
shown below, it corresponds to the one that naturally arises from the point
of view of Chern-Simons gauge theory. Let us expand the set of chord diagrams
to a new set in which trivalent vertices are allowed. This means that now
the lines in the interior of the circle can join a point on the circle to a
point on one of the internal lines. Fig. \ref{tresv} shows one of the new
allowed diagrams. Notice that the point that looks like a four-valent vertex
does not have any particular meaning. The new set of diagrams form a graded
vector space whose grading is half the total number of vertices (internal
trivalent vertices plus the previous vertices at the attachments of internal
lines to the circle). Bar-Natan showed \cite{barnatan} that the previous space
${\cal A}$ in (\ref{chordsp}) is equivalent to the new one after modding out
by the so-called  1T, AS, IHX and STU relations. The relation 1T is the
previous one shown in fig. \ref{fourt}. The relation AS is the statement that
the internal trivalent vertices are totally antisymmetric. Finally, the
relations STU and IHX are the ones shown in fig. \ref{stu}. Notice that in the
relation STU the curved line corresponds to a piece of the circle of the
diagram. The result proved by Bar-Natan in
\cite{barnatan} is simply:
\begin{equation}
{\cal A} = \Big\{ {\rm trivalent\,\,\, diagrams} \Big\} \,\, \Big/
\,\, {\rm (1T,\,\, AS,\,\, STU,\,\, IHX)}
\label{trisp}
\end{equation}
where ${\cal A}$ is the space (\ref{chordsp}).

The relations AS, STU and IHX are reminiscent of a Lie-algebra structure. If
one assigns totally antisymmetric structure constants $f_{abc}$ to the
internal trivalent vertices, and group generators $T_a$ to the vertices on the
circle, the STU relation is just the defining Lie-algebra relation,
\begin{equation}
f_{abc} T_c = T_a T_b - T_b T_a,
\label{lie}
\end {equation}
while the IHX relation corresponds to the Jacobi identity,
\begin{equation}
f_{abd} f_{dce} + f_{cad} f_{dbe} + f_{bcd} f_{dae} = 0.
\label{jacobi}
\end {equation}
We will find out below that the group factors associated to the perturbative
series expansion of the vev of a Wilson loop in Chern-Simons gauge theory
correspond precisely to the space ${\cal A}$ in (\ref{chordsp}) or
(\ref{trisp}) in its representation in terms of trivalent graphs. In that
context, the 1T relation is related to framing. In order to identify this
group factors with the space ${\cal A}$ one must consider the perturbative
series in a formal sense in which group factors are not particularized to a
definite group. The group factors must be regarded as objects which simply
satisfy relations (\ref{lie}) and (\ref{jacobi}) being $f_{abc}$ totally
antisymmetric. 

As follows from our discussion, singular knots play a central role in the
theory of Vassiliev invariants. We must now ask about their role in
Chern-Simons gauge theory. The first question to be addressed is about the
natural operators that should be associated to them. Wilson loops are related
to ordinary knots. What should be replacing the Wilson loop in the case of
singular knots while maintaining  Vassiliev resolution (\ref{vassreso})?
This question was answered in 1997 \cite{singular}. We will review here how
these operators for singular knots are obtained using quantum field theory
technics. The result will lead us to a proof of the theorem by Birman and Lin
discussed above, and to make direct contact with chord diagrams.

\begin{figure}
\centerline{\epsffile{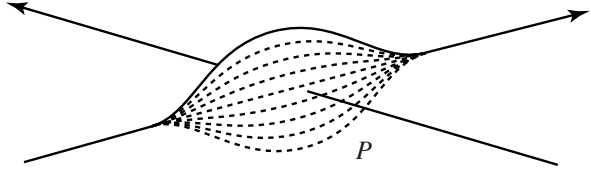}}
\caption{Family of paths with a intersection at the point $P$.}
\label{param}
\end{figure}

\begin{figure}
\centerline{\epsffile{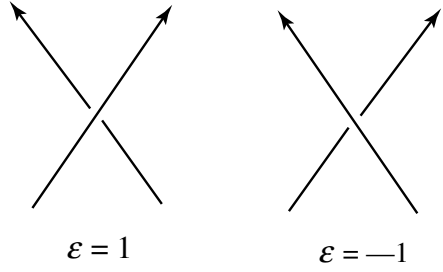}}
\caption{Signatures of an overcrossing and an undercrossing.}
\label{signat}
\end{figure}

We will begin studying the behavior of a difference of Wilson loops as the
one that enters in Vassiliev resolution (\ref{vassreso}). Let us consider a
family of smooth paths $\gamma_u$ parametrized by the continuous parameter
$u$, such that for $u=0$ the path $\gamma_0$ possesses a self-intersection at
some point $P$, \ie, for
$u=0$ it has a  double point. For
$u>0$ ($u<0$) the path presents an overcrossing (undercrossing) near the
point $P$. A family of paths with these features has been pictured in fig.
\ref{param}. The path $\gamma_0(v)$ has a double point at $v=s_1$ and
$v=t_1$ with
$s_1<t_1$. Paths $\gamma_u(v)$ with $u\neq 0$ are different from
$\gamma_0(v)$ only in the region in parameter space around $v=s_1$. The
derivative of $\gamma_u(v)$ with respect to $u$ is only non-zero in that
region. It vanishes away from $v=s_1$, in particular at $v=t_1$. In the
two resolutions of a double point an overcrossing
(undercrossing) corresponds to the one which leads to a crossing with
positive (negative) sign, as depicted in fig. \ref{signat}. Our goal is to
study the first derivative of the vev of the Wilson loop
$W_{\gamma_u}^R$ with respect  to the parameter $u$. Due to the topological
character of Chern-Simons gauge theory, one expects a step-function behaviour
for $\langle W_{\gamma_u}^R \rangle$ as a function of $u$ in the
neighborhood of $u=0$. This implies the presence of a delta function in its
derivative. As shown below, this is in fact what one finds. Our goal is
therefore to express
\begin{equation}
\langle W_{\gamma_+}^R \rangle  - \langle W_{\gamma_-}^R \rangle=
\int_{-\eta}^\eta du  {d\over du} \langle W_{\gamma_u}^R \rangle,
\label{martinezdos}
\end{equation}
where $\eta$ is some  positive small  real number, as the vev of some
operator.

To compute the integrand on the right hand side of
(\ref{martinezdos})  we will use a series of well known properties of the
Wilson loop and perform some formal manipulations in the integral functional
inherent to  vevs. Under a
deformation of its path, the Wilson loop behaves as:
\begin{equation}
{d\over d u} W_{\gamma_u}^R =
\oint dv\,  \gamma_u^{'\mu}(v) \dot\gamma_u^\nu(v)
   U_{\gamma_u}^R(s,v) F_{\mu\nu}(\gamma(v)) U_{\gamma_u}^R(v,s),
\label{variacion}
\end{equation}
where $U_{\gamma_u}^R(s,v)$ denotes a Wilson line which starts at $s$ and
ends at $v$, and  $F_{\mu\nu}$ is the curvature of the connection
$A_\mu$. In (\ref{variacion}) we have denoted derivatives with respect to the
path parameter by a dot and derivatives with respect to $u$ by a prime.
Recall that $\gamma_u^{'}(v)$ is only different from zero in the region in
parameter space around $v=s_1$.  Another important property of the Wilson
line is its behavior under a functional derivation with respect to the gauge
connection:
\begin{equation}
{\delta \over \delta A^a_\mu(x)} U_{\gamma_u}^R(s,t) =
\int_s^t dw\, \dot\gamma_u^\mu(w) \delta^{(3)}(x,\gamma_u(w))  
U_{\gamma_u}^R(s,w) T^a U_{\gamma_u}^R(w,t).
\label{derivada}
\end{equation}

Relations (\ref{variacion}) and (\ref{derivada}) are common to any gauge
theory. What makes Chern-Simons gauge theory special is that, after
performing a variation of the action respect to the gauge connection one
finds:
\begin{equation} {\delta \over \delta A_\mu^a(x)} S (A) = {k\over 8\pi}
\epsilon^{\mu\nu\rho} F_{\nu\rho}^a(x),
\label{fieldequation}
\end{equation}
\ie, the field equation involves just the field strength $F_{\mu\nu}$ and not
its derivative like in Yang-Mills theory. 

Taking into account (\ref{variacion}) and (\ref{fieldequation}), and
integrating by parts in connection space, one can write the integrand on the
right hand side of (\ref{martinezdos}) as:
\begin{equation} 
{d\over du} \langle W_{\gamma_u}^R \rangle=
{4\pi i \over k}\int [DA]
\ex^{i S(A)}
\oint dv\, \epsilon_{\mu\nu\rho} \gamma_u^{'\mu}(v)\dot\gamma_u^\nu(v)
 {\delta \over \delta A_\rho^a(\gamma_u(v))}
 U_{\gamma_u}^R(s,v) T^a U_{\gamma_u}^R(v,t).
\label{martin}
\end{equation} 
Using (\ref{derivada}) and disregarding some subtle contributions related to
framing \cite{singular} one finds:
\begin{equation} 
{d\over du} \langle W_{\gamma_u}^R \rangle=  {4\pi i \over
k}\delta(u) 
  \int [DA]
\ex^{i S(A)}  \tr\Big[ T^a U_{\gamma_u}^R(s_1,t_1) T^a
U_{\gamma_u}(t_1,s_1)\Big].
\label{martinazo}
\end{equation} 
As expected this expression involves a delta function in $u$. This proves in
turn that for continuous deformations of the path which do not involve
self-crossings the vev of a Wilson loop is invariant.

Using (\ref{martinezdos}) and the result (\ref{martinazo}) one can read the
operator which one must associate to a singular knot while maintaining 
Vassiliev resolution (\ref{vassreso}). Indeed, from 
(\ref{martinezdos}) and  (\ref{martinazo}) follows that
\begin{equation}
{4\pi i\over k} {\Bigg\langle} \tr\Big[T^a
 U_{\gamma}^R(s_1,t_1) T^a U_{\gamma}^R(t_1,s_1)\Big] {\Bigg\rangle}
=  \langle W_{\gamma_+}^R \rangle  - \langle W_{\gamma_-}^R \rangle,
\label{primero}
\end{equation}
and therefore $(4\pi i / k) \tr\Big[T^a
 U_{\gamma}^R(s_1,t_1) T^a U_{\gamma}^R(t_1,s_1)\Big]$ is the sought operator.
Notice that in (\ref{primero}) the subindex $u$ in the path has been
suppressed. One can easily show that this operator is gauge invariant
\cite{singular}. Its form is rather simple: split at the double point the
Wilson loop into two Wilson lines and insert two group generators.

The result obtained for a single double point generalizes easily to deal with
the general situation of $n$ double points. Let us consider a singular knot
 $K^n$ with $n$ double points, and let us assign to
each double point $i$ a triple $\tau_i=\{s_i,t_i,T^{a_i}\}$ where $s_i$ and
$t_i$, $s_i<t_i$, are the values of the $K^n$-parameter at the double
point, and $T^{a_i}$ is a group generator. The gauge-invariant operator
associated to the singular knot $K^n$ is:
\begin{eqnarray} && {\hskip-2cm} ({4\pi i \over k})^n \tr \Big[T^{\phi(w_1)}
U(w_1,w_2) T^{\phi(w_2)} U(w_2,w_3) T^{\phi(w_3)}\cdots \nonumber\\ &&
{\hskip3cm}
\cdots
U(w_{2n-1},w_{2n}) T^{\phi(w_{2n})} U(w_{2n},w_{1}) \Big],
\label{operador}
\end{eqnarray} 
where $\{w_i;i=1,\dots,2n\}$, $w_i<w_{i+1}$, is the set that
results from ordering the values $s_i$ and $t_i$, for $i=1,\dots,n$, and
$\phi$ is a map that assigns to each $w_i$ the group generator in the
triple to which it belongs.

The singular operators (\ref{operador}) lead to some immediate applications.
First, it is easily shown that they constitute a proof of the theorem by
Birman and Lin discussed above. This theorem was discussed at the beginning
of this section. It states that if in any polynomial knot invariant with
variable
$t$ one substitutes
$t\rightarrow
\ex^x$, and expand in powers of $x$, the coefficient of the power $x^n$ is a
Vassiliev invariant of order $n$ (see eq. (\ref{elteo})). A Vassiliev
invariant of order
$n$ vanishes for all singular knots with $n+1$ crossings. In the language of
Chern-Simons gauge theory the theorem by Birman and Lin can be rephrased,
simply stating that the $n^{\rm th}$-order term of the perturbative series
expansion of the vev of the Wilson loop associated
to a given knot is a Vassiliev invariant of order $n$. We will now prove
that this is so with the help of the operators (\ref{operador}).

The vevs of the operators (\ref{operador}) provide an
invariant for a singular knot with $n$ double points. This singular-knot
invariant can be expressed as a signed sum of $2^n$ invariants for
non-singular knots. These $2^n$ invariants are the perturbative series
expansion of the vev of the corresponding Wilson
loops. To show that the coefficient of order $n$ of these vevs is a Vassiliev
invariant or order $n$ is equivalent to proving that all the terms of order
less than $n$ vanish in the signed sum. But the signed sum is precisely the
vev of the operator  (\ref{operador}). Since the
perturbative series expansion of these operators starts at order $n$, the
proof of the theorem by Birman and Lin follows.

With the help of the operators (\ref{operador}) one makes direct contact with
chord diagrams. Let us consider the vevs of the operators
(\ref{operador}) at order zero. Their expressions are simply obtained by
setting $U=1$ in all the Wilson lines. Let us consider a singular knot
$K^n$  with $n$ double points, 
$s_i$, $t_i$, with $s_i<t_i$, for $i=1,\dots,n$, in parameter space. As
usual, with these data one constructs the triples: $\tau_i=\{s_i,t_i,
T^{a_i}\}$. At lowest order in perturbation theory the operators
(\ref{operador}) become the group factors,
\begin{equation}
v_n (K^n) = ({4\pi i\over k})^n \tr \Big[T^{\phi(w_1)} T^{\phi(w_2)}
T^{\phi(w_3)}\cdots T^{\phi(w_{2n})} \Big],
\nonumber\\
\label{wsystem}
\end{equation}
where the set $\{w_1,w_2,\dots,w_n\}$, with $w_j < w_{j+1}$,
is obtained by ordering the set $\{s_i,t_i; i=1,\dots,n\}$, and $\phi$ is the
induced map that assigns to each $w_j$ the index of the group generator
in the triple to which $w_j$ belongs. The indices entering
(\ref{wsystem}) are paired. This allows the association, to each operator
(\ref{wsystem}), of a diagram in which the $2n$ points are
distributed on a circle and the ones that possess the same value of
$\phi$ are joined by a line. The resulting diagrams are the
chord diagrams for singular knots (as the one shown in fig.
\ref{cdiag}). 

The quantities which result after the assignment of Lie-algebra data to
chord diagrams are called weight systems \cite{barnatan}. For each system one
chooses a group and a representation. As it will become clear in the next
section, they correspond to the group theory factors in the context of
Chern-Simons perturbation theory. 

\vfill
\newpage

\section{Vassiliev invariants and Chern-Simons perturbation theory}
\setcounter{equation}{0}

Vassiliev invariants of order
$m$ for singular knots with $m$ double points have been studied in the
previous section. From a mathematical point of view these are chord diagrams
and from a Chern-Simons gauge theory point of view they are group factors or
weight systems associated to chord diagrams. An important question that was
addressed from the mathematical side some years ago was whether or not a
weight system can be integrated to construct Vassiliev invariants for
non-singular knots. The answer to this question turned to be affirmative. From
the side of Chern-Simons gauge theory the positive answer to the question
follows simply from the existence of the perturbative series expansion. In
this section we will describe some of the basics of  Chern-Simons
perturbation theory.  In the next section we will discuss the
appearance of Vassiliev invariants from the integration of weight systems.

The perturbative study of Chern-Simons gauge theory started with the works by 
Guadagnini, Martellini and Mintchev \cite{gmm} and by Bar-Natan \cite{natan}.
These works dealt basically with the lowest non-trivial order in
perturbation theory.  To construct the perturbative series expansion of the
vev of an operator when dealing with a gauge theory one is forced to make a
gauge fixing. The first analysis of the Chern-Simons perturbation theory were
made in the covariant Landau gauge. Subsequent studies \cite{alla,alts} in
this gauge led to a framework linked to the theory of Vassiliev invariants,
which culminated with the configuration space integral approach
\cite{bt,dylan}. In the rest of this section we will describe the structure of
the perturbative series expansion of the vev of a Wilson loop in the covariant
Landau gauge. Chern-Simons perturbation theory in this gauge for the case of
more general three-manifolds has been considered in \cite{singer}.

For a perturbative analysis it is more convenient to rescale the field
entering the Chern-Simons action (\ref{valery}) in such a way that the
coupling constant appears in the three-vertex of the theory. Rescaling the
gauge field by
\begin{equation}
A_\mu \rightarrow gA_\mu,
\label{cinco}
\end{equation}
where $g=\sqrt{{4\pi\over k}}$, the Chern-Simons action (\ref{valery})
becomes,
\begin{equation}
S'(A_\mu)=\int \tr(A\wedge d A+{2\over 3}g A\wedge A\wedge
A).
\label{seiseq}
\end{equation}
This form of the action has the standard ${1\over 2}$ for the kinetic part since 
the trace of the  fundamental
representation is normalized as:
$\tr(T^aT^b)={1\over 2}\delta^{ab}$. Notice
that after the rescaling (\ref{cinco}) covariant derivatives contain the
coupling constant $g$. 

As stated above, to construct the perturbative series a
gauge fixing  must be performed.  We begin choosing a Lorentz gauge in which
\begin{equation}
\partial^\mu A_\mu=0.
\label{cincoprime}
\end{equation}
The standard Faddeev-Popov construction leads us to
consider the following gauge-fixed functional  integral:
\begin{equation}
 \int[DA_\mu
DcD\bar cD\phi] \ex^{iS'(A_\mu)+
iS_{\scriptstyle {\rm gf}}(A_\mu,c,\bar
c,\phi)},
\label{presiete}
\end{equation}
where
\begin{equation}
S_{\scriptstyle {\rm gf}}(B_\mu,c,\bar c,\phi)=
\int {\rm d}^3 x \tr( 2\bar c
\partial_\mu D^\mu c 
-2 \phi \partial_\mu A^\mu - \lambda  \phi^2).
\label{preocho}
\end{equation}
In this action $\phi$ is a Lagrange
multiplier which imposes the gauge condition,  $c$ and $\bar c$ are
anticommuting Faddeev-Popov ghosts, and $\lambda$ is  a gauge-fixing
parameter. The derivative in $D_\mu c$ is a covariant derivative. The
functional integral in (\ref{presiete}) must be done over all
$A_\mu$ configurations and not only over gauge orbits.
As a result of the gauge fixing, the  exponent in (\ref{presiete}) is
invariant under the corresponding BRST transformations.

The field $\phi$ can be integrated out by performing a Gaussian integration in
(\ref{presiete}). One finds, up to an irrelevant multiplicative factor, that
the functional integral (\ref{presiete}) becomes:
\begin{equation}
\int[DA_\mu
DcD\bar c] \ex^{iI(A_\mu,c,\bar c)},
\label{nueve}
\end{equation}
where
\begin{equation}
I(A_\mu,c,\bar c)=
\int \tr\big[\varepsilon^{\mu\nu\rho} (A_\mu\partial_\nu A_\rho
+{2\over 3}g A_\mu A_\nu A_\rho) -{1\over \lambda}A_\mu\partial^\mu\partial^\nu 
A_\nu + 2\bar c \partial_\mu D^\mu c  \big].
\label{ocho}
\end{equation}
In order to compute the vevs of operators involving Wilson loops like in
(\ref{encarna}) one must integrate these operators using (\ref{nueve}).
After expanding the path-ordered products in the Wilson loops one ends
computing the functional integral over products of gauge fields. These are
basically the correlation functions of the theory and are computed using
standard quantum field theory methods, which lead to a set of computational
rules called Feynman rules. These rules provide terms which are
organized in even powers of $g$. They dictate that at order
$g^{2m}$ two kinds of diagrams must be taken into account. The first  group
involves  oriented trivalent graphs of the kind introduced in the previous
section with $2m$ vertices and with the following assignments: for each
internal line (gauge propagator),
\begin{equation}
{i\over 4\pi} \delta_{ab}\epsilon^{\mu\nu\rho}{(x-y)_\rho
\over |x-y|^3},
\label{nueveuno}
\end{equation}
for each internal vertex,
\begin{equation}
-i g f_{abc} \epsilon_{\mu\nu\rho} \int d^3x ,
\label{nuevedos}
\end{equation}
and for each vertex on the circle,
\begin{equation}
g (T^a)_i^j.
\label{nuevetres}
\end{equation}

The second set of diagrams involves propagators and vertices related to ghost
fields. Contrary to gauge field propagators, dashed lines are used for ghost
field propagators. The diagrams of the second set are also 
trivalent diagrams but they must contain at least one dashed line. Dashed
lines can not be attached to the circle. They are always attached to internal
lines via the ghost-gauge field-ghost trivalent vertex. We will not reproduce
here the form of the ghost-related ingredients of the Feynman rules. We will
describe, however, what is their effect.

In writing the quantities associated to the Feynman rules (\ref{nueveuno}),
(\ref{nuevedos}) and (\ref{nuevetres})  the limit
$\lambda\rightarrow 0$ has been taken. This is known as the Landau gauge. It
simplifies the calculations since in this case there are not infrared
divergences.

After applying the Feynman rules one finds that the
perturbative series contains divergent
terms.  Since the coupling constant $g$ is dimensionless, the theory is,
however, renormalizable. Power counting analysis \cite{shift} shows that
actually the theory is superrenormalible. To deal with the divergent terms, the
theory must be regularized. One can use a variety of regularizations. For most
of the regularizations which have been studied it turns out that the theory is
in fact finite, \ie, after removing the cutoff introduced in the regularization
the resulting expressions are finite. Of course, one is free to choose an
arbitrary scheme using a finite renormalization. The value of $k$ would be
different in each scheme if its fixed from some ``standard data". In other
words, the value of $k$ in a given scheme should be determined stating,
for example, that the value of the vev of the Wilson loop for the trefoil knot
divided by the partition function is a fixed quantity. Then, though working in
different schemes, the computations of vevs for any other knot would agree.

Many regularizations have ben studied in the last ten years. There is a
particular subset of them \cite{semen,shift,cmartin,asorey} that naturally
leads to a scheme in which a shift in the coupling constant $k$ occurs. In
these regularizations the higher-loop contributions to the two- and three-point
functions add up to a shift in
$k$ which is precisely the dual of the Coxeter number of the gauge group.
This is rather remarkable because it is the same shift that appears in
non-perturbative studies of Chern-Simons gauge theory in connection with its
associated two-dimensional conformal field theory \cite{csgt}. Why this is so
is not still well understood.

In our perturbative analysis we are going to assume that higher-loop
corrections to two- and three-point functions just account for the shift in $k$
and therefore we will not consider them in the expansion. We will deal with
the rest of the diagrams. The scheme is chosen so that we do not make any
finite renormalization. This scheme is as good as any other but is the best to
compare perturbative results to non-perturbative ones. As we will
discuss in the next sections, in non-covariant gauges we will make a choice
in which there is no shift and then, to compare to non-perturbative
calculations, we must make a finite renormalization.

Before writing the form of the full perturbative series we must deal with
another important subtlety. If one computes the first order contribution to
the perturbative series expansion of the vev of a Wilson loop one finds that
the resulting quantity is not a topological invariant. In the gauge fixing of
the theory we have introduced a metric dependence that could lead to
quantities which are not topological. This first order contribution is just a
manifestation of it. Fortunately, only in this term, and in its propagation in
higher order contributions, topological invariance is lost. The rest of the
perturbative series expansion is truly topological. Thus, although vevs
are not topological invariant quantities they fail to be
so in a controllable way. The non-topological terms factorize and multiply a
term which is topological. In non-perturbative studies one finds a related
problem which is the need of the introduction of a framing for the knot under
consideration \cite{csgt}. In other words, instead of knots one must consider
framed knots, or knots with a normal vector field assigned which defines a
framing characterized by an integer. This integer is the number of times that
the normal vector field winds around the knot. A framing can also be
introduced in the perturbative approach so that the perturbative result
coincides with the non-perturbative one. Let us describe how this is done.

The first order contribution to the vev of a Wilson loop has the form:
\begin{eqnarray}
 && g^2 \tr(T^aT^b) \oint d x^\mu \int^x d y^\nu {i\over 4\pi} \delta_{ab}
\epsilon^{\mu\nu\rho}{ (x-y)_\rho \over |x-y|^3} \nonumber \\
&=&  g^2 {1\over 2} \tr(T^aT^b) \oint d x^\mu \oint d y^\nu {i\over 4\pi}
\delta_{ab} \epsilon^{\mu\nu\rho}{ (x-y)_\rho \over |x-y|^3}.
\label{diez}
\end{eqnarray}
This expression is finite but depends on the shape of the knot. Let us
introduce a framing and, together with the original knot,  a
companion knot located at the end point of the normal vector which defines
the framing. If one replaces one of the original paths entering (\ref{diez})
by the path associated to the companion knot (which will be denoted by a
prime), one finds Gauss formula for the linking number:
\begin{equation}
 g^2 {1\over 2} \tr(T^aT^b) \oint d x^\mu \oint' d y^\nu {i\over 4\pi}
\delta_{ab} \epsilon^{\mu\nu\rho}{ (x-y)_\rho \over |x-y|^3} = l. 
\label{once}
\end{equation}
In this expression $l$ is an integer that counts the number of times that the
vector associated to the framing winds the knot. The kind of point splitting
associated to the framing leads to a perturbative result that agrees with the
non-perturbative analysis. Either if we introduce the framing or we leave a
non-topological term we obtain a good perturbative series expansion. The
corresponding term factorizes. Thus, to deal with one or the other is a matter
of taste but, as in the case of the shift, agreement with the non-perturbative
analysis induces the approach based on the framing. We will 
follow this choice. The term that factorizes has the form:
\begin{equation}
\exp\big(2\pi i l h_R  \big), \,\,\,\,\,\,\,
h_R = {1\over k+g^\vee}\tr_R(T^a T^a).
\label{doce}
\end{equation}
The quantity $h_R $ can be identified with the conformal weight for the
representation $R$ of the associated conformal field theory. The Feynman
diagrams that lead to the framing factor are those diagrams which contain
isolated chords once a canonical basis from group factors has been chosen.
Canonical basis play a prominent role in our discussion but before defining
them let us first take a look at the general form of the perturbative series.

Once we have dealt with the issues regarding the shift and the framing we
are in the position to analyze the perturbative series expansion of the vev
of a Wilson loop. The Feynman rules
(\ref{nueveuno}), (\ref{nuevedos}) and (\ref{nuevetres}) allow to split the
contributions to each order in two factors: a geometrical factor which
includes all the space dependence, and  a group factor which includes all the
group theoretical dependence. The general form is:
\begin{equation}
\langle W^R_K \rangle = \hbox{\rm dim}\, R \sum_{i=0}^\infty
\sum_{j=1}^{d_i}\alpha_{ij}(K) r_{ij}(R) x^i,
\label{expansion}
\end{equation}
where,
\begin{equation}
x={2\pi i\over k}=ig^2/2 
\label{laequis}
\end{equation}
is the expansion parameter, 
$\hbox{\rm dim}\, R$ is the dimension of the representation $R$,  
$\alpha_{0,1}=r_{0,1}=1$, and $d_0=1$. Notice that higher-loop corrections
to the two- and three-point functions must not be included in (\ref{expansion})
so this series should correspond to the non-perturbative result without the
shift. The factors $\alpha_{ij}(K)$ and
$r_{ij}(R)$ appearing at each order $i$ incorporate all the dependence
dictated from the Feynman rules apart from the dependence on  the
coupling constant, which is contained in $x$. 
Of these two factors, in the $r_{ij}(R)$ all
the group-theoretical dependence is collected. These are the 
group factors mentioned above. The rest is contained in  the $\alpha_{ij}(K)$
or geometrical factors. They have the form of integrals over the Wilson loop
corresponding to the knot $K$ of products of propagators, as dictated by the
Feynman rules. The first index in $\alpha_{ij}(K)$ denotes the order in the
expansion and the second index labels the different geometrical factors which
can contribute at the given order. Similarly, $r_{ij}(R)$ stands for the
independent group structures which appear at order
$i$, which are also dictated by the Feynman rules. The object
$d_i$ in (\ref{expansion}) is the dimension of the space of
invariants at a given order. In our approach denotes the number of
independent group structures which appear at that order.  Notice that while
the geometrical factors $\alpha_{ij}(K)$ are knot dependent but group and
representation independent, the group factors are group and representation
dependent but knot independent.

\begin{figure}
\centerline{\epsffile{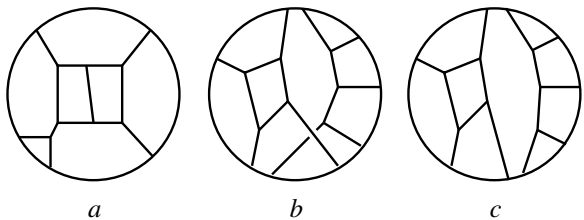}}
\caption{Examples of trivalent diagrams: {\it a} is connected while {\it b}
and {\it c} are disconnected.}
\label{ejemplo}
\end{figure}

Among the basis of group factors which can be chosen in the perturbative
series (\ref{expansion}) there is a special class which turns out to be very
useful. We will call it the class of canonical basis. Notice that to each
group factor one can associate a trivalent diagram of the kind entering
(\ref{trisp}) if one considers the space of all semi-simple Lie groups (as
that is the most general case for which the structure constants can be chosen
totally antisymmetric). We will restrict to this case in what follows. 

In order to introduce the concept of canonical basis we must first deal with a
sort of classification of trivalent diagrams. A trivalent
diagram will be called connected diagram if it is  possible to go
from one  propagator (or internal line) to another without ever having to go
through  the external circle. The diagram will be called disconnected diagram
if that is not possible. In  this second case we say that the diagram has
subdiagrams, which are the  connected components of the whole diagram.
Two  subdiagrams  are non-overlapping if we can move along
the external circle meeting all the  legs of one subdiagram first, and all the
legs of  the other  second. By legs we mean internal lines 
attached to the external circle. If it is impossible to do that, the
subdiagrams are overlapping. In fig. \ref{ejemplo} the diagram {\it a} is
connected while the diagrams {\it b} and {\it c} are disconnected. Of these
last two, diagram {\it b} contains subdiagrams which are overlapping while 
{\it c} does not.

\begin{figure}
\centerline{\epsffile{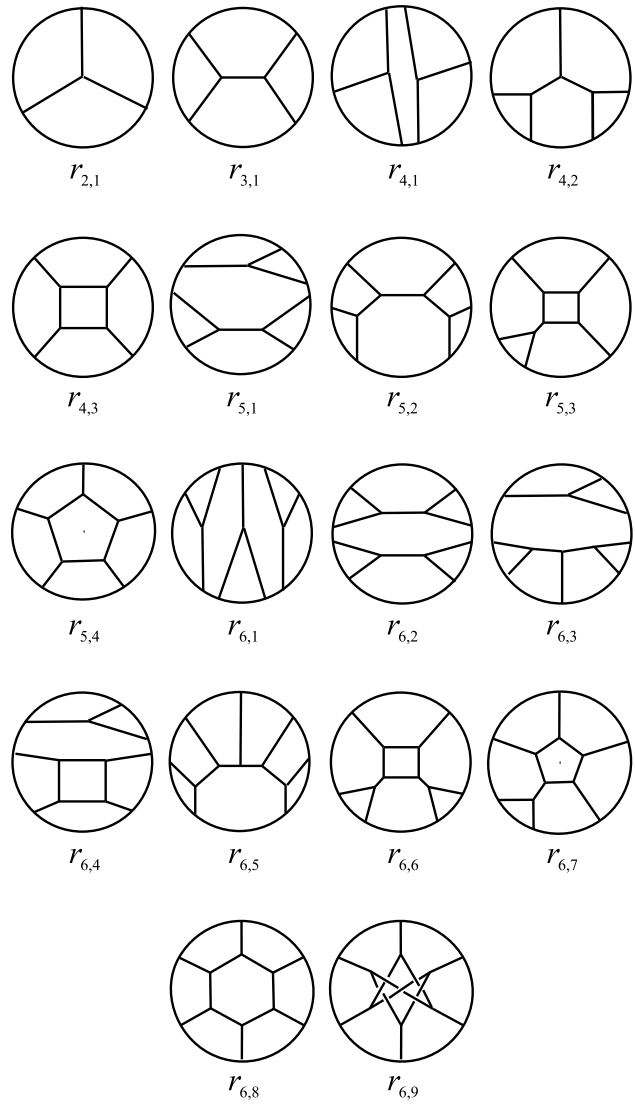}}
\caption{Choice of a canonical basis up to order 6.}
\label{canonical}
\end{figure}

In the expansion (\ref{expansion}) there are many possible choices of
independent groups factors $r_{ij}(R)$. Given all trivalent diagrams
contributing to a given order in perturbation theory some of the resulting 
group factors might be related due to the Lie-algebra relations AS, STU and
IHX as in (\ref{trisp}). The group factors entering 
(\ref{expansion}) are chosen to be associated to diagrams that constitute a
basis. Its elements are the $r_{ij}(R)$ in  (\ref{expansion}). Of course, many
choices are possible. In order to introduce our preferred choice let us first
notice that due the STU relations one can always choose
a basis such that the $r_{ij}(R)$ come from 
connected diagrams, or products of connected diagrams. That is, if there are
subdiagrams, they can be chosen so that they do not overlap. The value of such 
an $r_{ij}(R)$ is the product of the values of its subdiagrams.
This last statement follows
from the fact that the part of the group factor associated to one of the
subdiagrams is a diagonal matrix.
Thus one can choose a basis built of connected trivalent diagrams and
products of connected non-overlapping subdiagrams. A basis with this feature
will be called a canonical basis. In fig. \ref{canonical} a choice of
canonical basis up to order six is shown.

We will denote by $r_{ij}^c(R)$ the group factors associated to a canonical
basis, and $\alpha_{ij}^c(K)$ the corresponding geometrical factors. It
was shown in \cite{factor} that then the perturbative series expansion
(\ref{expansion}) can be written as:
\begin{equation}
\langle W^R_K \rangle = \hbox{\rm dim}\, R \,\exp\left\{ \sum_{i=1}^\infty
\sum_{j=1}^{\hat d_i} \alpha_{ij}^c(K)\,r_{ij}^c(R)\,x^i\right\},
\label{expansiondos}
\end{equation}
where $\hat{d}_i$ stands for the number of connected elements in the
canonical basis at order $i$. Notice that $\alpha_{ij}^c(K)$ do not
correspond uniquely to connected Feynman diagrams. The result
(\ref{expansiondos}) is known as the factorization theorem. Actually, it holds
for arbitrary gauges,  not only in the covariant Landau gauge as one could
conclude from our discussion. The dimensions $\hat d_i$ are precisely the
ones introduced for the connected part of the space of chord diagrams
(\ref{chordsp}). Their values up to order 12 are the ones contained in the
table  shown after (\ref{chordsp}). The geometrical factors 
$\alpha_{ij}^c(K)$ are a selected set of Vassiliev invariants. They are
called primitive Vassiliev invariants. If they are known, the values
of the whole set of Vassiliev invariants follow. These primitive Vassiliev
invariants have been computed for general classes
of knots as torus knots \cite{torus,simon} up to order six.

The contribution at first order in (\ref{expansiondos}) is precisely the
framing factor (\ref{doce}). Thus, the factorization theorem
(\ref{expansiondos}) shows also its factorization. The rest of the terms in
the exponent of (\ref{expansiondos}) are knot invariants. The series
contained in that exponent was analyzed by Bott and Taubes \cite{bt} in their
work on the configuration space for Vassiliev invariants. They showed that the
integral expression entering the geometrical factors $\alpha_{ij}^c(K)$ are
convergent \cite{bt,dylan}. Further work on the subject has led to a proof of
their invariance
\cite{alts,yang}.

The explicit expression for the integrals entering in the second order
contribution was first presented in  \cite{gmm}. It was later analyzed in
detail by Bar-Natan
\cite{natan}. It takes the form:
\begin{eqnarray}
\alpha_{21}^c(K)&=&\frac{1}{4}\oint dx^\mu \int^x d y^\nu \int^y dz^\rho
\int^z dw^\sigma \Delta_{\mu\rho}(x-z) \Delta_{\nu\sigma}(y-w) \nonumber \\
&-&\frac{1}{16}\oint dx^\mu \int^x d y^\nu \int^y d z^\rho
\int d^3\omega\, v_{\mu\nu\rho}(x,y,z;\omega),
\label{sara}
\end{eqnarray}
where,
\begin{equation}
\Delta_{\mu\nu}(x-y) = \frac{1}{\pi}\epsilon_{\mu\rho\nu}
\frac{(x-y)^\rho}{|x-y|^3},
\label{adela}
\end{equation}
and,
\begin{equation}
v_{\mu\nu\rho}(x,y,z;\omega)=\Delta_{\mu\sigma_1}(x-\omega)
\Delta_{\nu\sigma_2}(y-\omega)\Delta_{\rho\sigma_3}(z-\omega)
\epsilon^{\sigma_1 \sigma_2 \sigma_3}.
\label{adelaida}
\end{equation}
This invariant turns out to be the total twist of the knot and coincides mod 2
with the Arf invariant. The integral expression for the order three
invariant, 
$\alpha_{31}^c(K)$, associated to the group factor $r_{31}$ shown in fig.
\ref{canonical} was first presented in  \cite{alla}. We reproduce it here for
completeness:
\begin{eqnarray}
\alpha_{31}^c(K)&&={1\over 8} 
\oint \,{dx_1^{\mu_1}} 
\int^{x_1} \,{dx_2^{\mu_2}} 
\int^{x_2} \,{dx_3^{\mu_3}} 
\int^{x_3} \,{dx_4^{\mu_4}} 
\int^{x_4} \,{dx_5^{\mu_5}} 
\int^{x_5} \,{dx_6^{\mu_6}} \nonumber \\
 \big[ &&
\Delta_{\mu_1\mu_4}(x_1-x_4)\Delta_{\mu_2\mu_6}(x_2-x_6)
\Delta_{\mu_3\mu_5}(x_3-x_5) \nonumber \\ && +
\Delta_{\mu_1\mu_3}(x_1-x_3)\Delta_{\mu_2\mu_5}(x_2-x_5)
\Delta_{\mu_4\mu_6}(x_4-x_6)  \nonumber \\ &&
+\Delta_{\mu_1\mu_5}(x_1-x_5)\Delta_{\mu_2\mu_4}(x_2-x_4)
\Delta_{\mu_3\mu_6}(x_3-x_6)
\big] \nonumber \\
+{1\over 4}&&
\oint \,{dx_1^{\mu_1}} 
\int^{x_1} \,{dx_2^{\mu_2}} 
\int^{x_2} \,{dx_3^{\mu_3}} 
\int^{x_3} \,{dx_4^{\mu_4}} 
\int^{x_4} \,{dx_5^{\mu_5}} 
\int^{x_5} \,{dx_6^{\mu_6}} \nonumber \\
 \big[ &&
\Delta_{\mu_1\mu_4}(x_1-x_4)\Delta_{\mu_2\mu_5}(x_2-x_5)
\Delta_{\mu_3\mu_6}(x_3-x_6) 
\big] \nonumber \\
-{1\over 32}&&
\oint \,{dx_1^{\mu_1}} 
\int^{x_1} \,{dx_2^{\mu_2}} 
\int^{x_2} \,{dx_3^{\mu_3}} 
\int^{x_3} \,{dx_4^{\mu_4}} 
\int^{x_4} \,{dx_5^{\mu_5}}
\int d^3y  \\
 \big[ &&
\Delta_{\mu_1\nu_1}(x_1-y)\Delta_{\mu_2\mu_5}(x_2-x_5)
\Delta_{\mu_3\nu_3}(x_3-y)\Delta_{\mu_4\nu_4}(x_4-y)
\epsilon^{\nu_1\nu_3\nu_4} \nonumber \\
&&+ \Delta_{\mu_1\mu_3}(x_1-x_3)\Delta_{\mu_2\nu_2}(x_2-y)
\Delta_{\mu_4\nu_4}(x_4-y)\Delta_{\mu_5\nu_5}(x_5-y)
\epsilon^{\nu_2\nu_4\nu_5} \nonumber \\
&&+\Delta_{\mu_1\nu_1}(x_1-y)\Delta_{\mu_2\mu_4}(x_2-x_4)
\Delta_{\mu_3\nu_3}(x_3-y)\Delta_{\mu_5\nu_5}(x_5-y)
\epsilon^{\nu_1\nu_3\nu_5} \nonumber \\
&&+ \Delta_{\mu_1\nu_1}(x_1-y)\Delta_{\mu_2\nu_2}(x_2-y)
\Delta_{\mu_3\mu_5}(x_3-x_5)\Delta_{\mu_4\nu_5}(x_4-y)
\epsilon^{\nu_1\nu_2\nu_4} \nonumber \\
&&+\Delta_{\mu_1\mu_4}(x_1-x_4)\Delta_{\mu_2\nu_2}(x_2-y)
\Delta_{\mu_3\nu_3}(x_3-y)\Delta_{\mu_5\nu_5}(x_5-y)
\epsilon^{\nu_2\nu_3\nu_5} \big] \nonumber \\
+{1\over 128}&&
\oint \,{dx_1^{\mu_1}} 
\int^{x_1} \,{dx_2^{\mu_2}} 
\int^{x_2} \,{dx_3^{\mu_3}} 
\int^{x_3} \,{dx_4^{\mu_4}} 
\int d^3y \int d^3z\nonumber \\
\big[ &&
\Delta_{\mu_1\nu_1}(x_1-y)\Delta_{\mu_2\sigma_2}(x_2-z)
\Delta_{\mu_3\sigma_3}(x_3-z)\Delta_{\mu_4\nu_4}(x_4-y)
\Delta_{\alpha\beta}(y-z)\epsilon^{\nu_1\alpha\nu_4}
\epsilon^{\sigma_3\beta\sigma_2}\nonumber \\
&&+\Delta_{\mu_1\nu_1}(x_1-y)\Delta_{\mu_2\nu_2}(x_2-y)
\Delta_{\mu_3\sigma_3}(x_3-z)\Delta_{\mu_4\sigma_4}(x_4-z)
\Delta_{\alpha\beta}(y-z)\epsilon^{\nu_2\alpha\nu_1}
\epsilon^{\sigma_4\beta\sigma_3} \big] \nonumber
\label{ordentres}
\end{eqnarray}

Properties of the  primitive Vassiliev invariants $\alpha_{21}^c(K)$ and
$\alpha_{31}^c(K)$ have been studied in
\cite{alemanes}. In these works the integral
expressions for $\alpha_{21}^c(K)$ and $\alpha_{31}^c(K)$ were studied in the
flat-knot limit and combinatorial expressions of the kind that will be
presented below from the study in the temporal gauge were obtained.

\vfill
\newpage

\section{Light-cone gauge and the Kontsevich integral}
\setcounter{equation}{0}

In the previous section, the perturbative series expansion of
the vev of a Wilson loop in the covariant Landau gauge has been analyzed. A
similar analysis can be carried out for the operators associated to singular
knots which were introduced in sect. 3. The lowest order contributions in
such an expansion are the group factors corresponding to chord diagrams.
These group factors form a weight system.  Before the
development of operators for singular knots the following question was raised:
could weight systems on chord diagrams be integrated to obtain invariants for
non-singular knots? Kontsevich answered this question affirmatively in 1993.
He showed that a weight system on chord diagrams of order $m$ determines a
unique Vassiliev invariant on non-singular knots.

Kontsevich theorem is constructive in the sense that it provides an explicit
expression for the Vassiliev invariant for non-singular knots. This
expression is known as the Kontsevich integral. We will provide its explicit
form below, after deriving it from the quantum field theory side. In
the context of Chern-Simons gauge theory Kontsevich theorem is contained in
our construction of operators for singular knots. One should think of the
first order contribution of the vev of a singular knot as the weight system,
and of the full perturbative series for vevs of Wilson loops as the
integration of the weight system. Thus, the perturbative series expansion
of the vev of a Wilson loop should correspond to the Kontsevich integral. This
does not seem to be the case as the configuration space integrals found in the
previous section have a different form than the integrals contained in
Kontsevich theorem. This fact poses a puzzle which, however, is solved
very simply.

Gauge theories can be studied in different gauges. The analysis carried out
in the previous sections dealt with the covariant Landau gauge but many
others could have been used. It turns out that the perturbative series
which results in the non-covariant light-cone gauge leads to the Kontsevich
integral. To show this is the main goal of this section. Vevs of gauge
invariant operators are independent of the gauge chosen and therefore the
expressions obtained in the light-cone gauge should be equivalent to the ones
obtained in any other gauge. In this sense the analysis of operators for
singular knots in Chern-simons gauge theory constituted a general proof, or
gauge independent proof, of Kontsevich theorem.  The configuration space
integrals of the previous section can equally be regarded as an integration of
weight systems on chord diagrams. Actually, very recently, it has been shown
from the mathematical side \cite{poir} that both series of integrals are
equivalent. A  ratification of the gauge independence of gauge invariant
operators mentioned above.

In order to deal with non-covariant gauges let us introduce a unit constant
vector $n$. The gauge-fixing conditions of axial type which we will
discuss in this and in the next section correspond to the choice:
\begin{equation}
 n^{\mu} A_{\mu} = 0.
\label{light}
\end{equation} 
In the case of the light-cone gauge the unit vector $n$ satisfies the
condition $n^2=0$. The gauge fixed action of the theory is built
following the standard Faddeev-Popov construction which leads to a functional
integral as in (\ref{presiete}) where now:
\begin{equation}
S_{\scriptstyle {\rm gf}}(B_\mu,c,\bar c,\phi)=
\int {\rm d}^3 x \tr ( \phi
n^{\mu} A_{\mu} + b n^{\mu} D_{\mu} c +\lambda \phi^2).
\label{preochodos}
\end{equation}
Again, $\phi$ is a Lagrange
multiplier which imposes the gauge condition,  $c$ and $\bar c$ are
anticommuting Faddeev-Popov ghosts, and $\lambda$ is  a gauge-fixing
parameter. 

The higher-loop analysis in axial-type gauges is a very delicate
issue. It  requires to take into consideration some specific
prescription to regulate unphysical poles. Fortunately, this analysis has
been done for the case of Chern-Simons gauge theory in \cite{leibbrandt}.
In these works it is shown that a regulator can be chosen so that the effect
of higher-loop contributions for two- and three-point functions is a shift of
the parameter $k$ as the one discussed in the previous section in the
covariant Landau gauge. Though strictly speaking this has been proved at one
loop, it is believed that, as in the case of covariant gauges, it holds at any
order in higher loops. We will assume that higher-loop contributions to the
two- and the three-point functions account for the shift of the parameter
$k$. Notice that the gauge condition (\ref{light}) does not fix the gauge
completely. One still has gauge invariance under gauge transformations in
the direction normal to $n$. The presence of this residual gauge
invariance is a source of problems which, as it will be discussed below, are
not yet solved. Another study of light-cone perturbation theory can be found
in \cite{sorella}.

We will work in the limit $\lambda \rightarrow 0$. The theory greatly
simplifies in this case. The propagator for the gauge field in momentum space
acquires the simple form:
\begin{equation}
\delta_{ab} {\lambda \over (np)^2}(p_\mu p_\nu - {i\over \lambda} (np)
\epsilon_{\mu\nu\rho} n^\rho )  \rightarrow
-i \epsilon_{\mu\nu\rho} {n^\rho \over n p} \delta_{ab}.
\label{nprop}
\end{equation}
This propagator is orthogonal to the $n$-direction. This implies that there
is no coupling to the ghosts fields and, furthermore, that there is no
gauge field self-coupling. Thus there is only a Feynman rule to be taken into
account to compute the vevs of operators: the one associated to the gauge
field propagator. The group factors that remain in this case correspond just
to chord diagrams. The fact that in this gauge  no group factors with
trivalent vertices have to be taken into account is a quantum field theory
ratification of Bar-Natan theorem among the equivalence of the two
representations, (\ref{chordsp}) and (\ref{trisp}), of the space ${\cal A}$.

Non-covariant gauges share the problem of the presence of unphysical poles in
their propagators \cite{leibrew}. This is manifest in (\ref{nprop}). Several
prescriptions have been proposed to avoid these unphysical poles. Usually, a
prescription is chosen so that some particular properties of the correlation
functions are fulfilled. In the light-cone gauge there is a natural
prescription which is motivated by the simple form that (\ref{nprop}) takes
in coordinate space after performing a Wick rotation. As it will be shown
below, this prescription leads to Kontsevich integral.

In the light-cone gauge it is convenient to  introduce
light-cone coordinates. Let us choose the unit vector $n$ as $(0,1,-1)$.
The new coordinates are:
\begin{equation} 
x^+ = x^1 + x^2,   \,\,\, \,\,\, x^- = x^1-x^2,
\label{coor}
\end{equation} 
and the new light-cone components for the gauge connection become:
\begin{equation} A_+ = A_1 + A_2,   \,\,\, \,\,\, A_- = A_1-A_2.
\label{lcgc}
\end{equation} 
The form of the propagator (\ref{nprop}) in coordinate space is more
conveniently expressed after performing a Wick rotation to Euclidean space. 
A point in Euclidean space will be denoted as
$(t,z)$, where $z=x^1+i x^2$. After introducing $A_z = A_1 + i A_2$ and
$A_{\bar z} = A_1 - i A_2$ one finds that there exist a prescription to deal
with the unphysical pole in (\ref{nprop}) which leads to:
\begin{eqnarray}
\langle A_{\bar z}^a(x) A_m^b(x') \rangle &=& 0, \nonumber
\\  \langle A^a_m(x) A^b_n(x') \rangle &=&
 \delta^{ab} \epsilon_{mn} { 1 \over 2\pi i}  { \delta ( t - t'  )
\over z - z' },
\label{maspropa}
\end{eqnarray}
with $m,n = \{ 0,z \}$, and $\epsilon_{mn}$ is antisymmetric with 
$\epsilon_{0z}=1$. This is the gauge-field propagator that we will use in our
analysis of the perturbative expansion of the vev of a Wilson loop.

\begin{figure}
\centerline{\epsffile{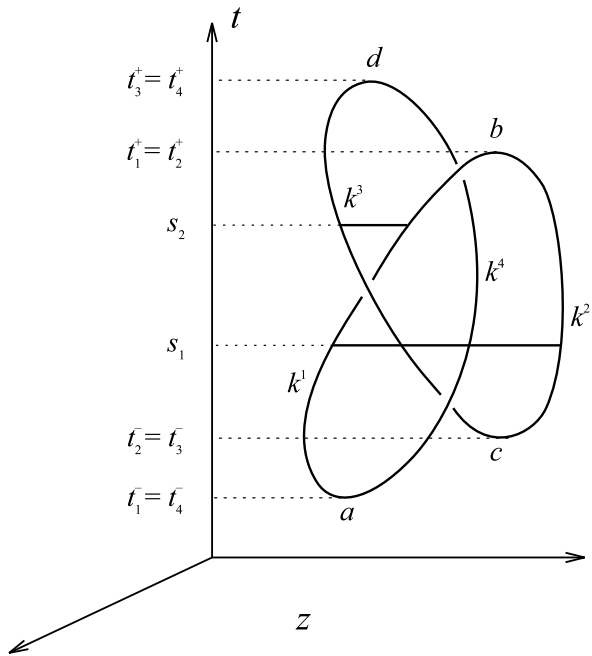}}
\caption{Example of a Morse knot.}
\label{morse}
\end{figure}

Before entering into the analysis of the perturbative series expansion
corresponding to the vev a Wilson loop, we must discuss the potential problems
that might be encountered because of the particular form of the
gauge-field propagator (\ref{maspropa}). This propagator is singular when
its two end-points coincide. Actually, it is particularly singular in this
situation because both the numerator and the denominator lead to
divergences. In the light-cone gauge we have two special kinds of
singularities, there may be situations in which only one, the numerator
or the denominator, leads to a divergence.  In order to avoid
singularities from the numerator one is forced to avoid paths with
sections in which $t$, the first component of a generic point $(t,z)$ 
is constant. This constraint implies that one must consider  Morse knots. A
Morse knot is a knot in which $t$ is a Morse function on it. A Morse
knot is characterized by $2n$ extrema, half of them maxima, and the other
half minima. An example of a Morse knot is depicted in fig.
\ref{morse}. 

Another potential problem due to the structure of (\ref{maspropa}) comes
from situations in which the two end-points of the propagator are close to
one of the extrema of a Morse knot since  the denominator then vanishes.
To solve this problem one introduces a regularization procedure based
on the introduction of a framing. The resulting invariants
will correspond to invariants of framed knots. As in the case of the Landau
gauge one finds \cite{lcone} that a term like (\ref{doce}) factorizes. One is
left then with a perturbative series expansion with group factors which
correspond to chord diagrams without isolated chords. The ones with isolated
chords just contribute to the framing factor.

In order to obtain the form of the perturbative series of the vev of a Wilson
loop we begin by writing  all the contributions at a given order
$m$. To carry this out we must consider all possible ways of connecting  $2m$
points on the Morse knot by $m$ propagators, taking into account the
regularization described above, \ie, with one
point of the propagator attached to $K$ and the other to its companion knot
$K_\varepsilon$, and then path-order integrating. The path-order
integration can be split into a sum such that in each term enters a
path-ordered integration along $2m$ curves among the set $k^i$, 
$k^i_\varepsilon$, $i=1,\dots,2n$. This set of curves builds  the Morse
knot under consideration.  A given term in this sum might contain
propagators joining  $k^i$ and $k^i_\varepsilon$. In this case one must
introduce a factor $1/2$. The contributions coming from propagators joining
$k^i$ and $k^j_\varepsilon$, with $i\neq j$, have also a factor $1/2$ since
double counting occurs. Accordingly, propagators joining different curves must
be replaced by $1/2$ the sum of their two possible choices of attaching their
end-points. 

\begin{figure}
\centerline{\epsffile{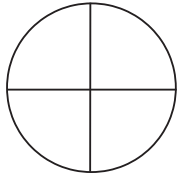}}
\caption{Group factor at second order.}
\label{cruz}
\end{figure}

To each rearrangement of the $m$ propagators corresponds a group factor.
These are easily obtained just using the rule  (\ref{nuevetres}). To fix ideas
we will present in detail the second-order contribution, $m=2$, for a
particular group factor, the one shown in fig. \ref{cruz}.  This contribution
is of the form
\begin{eqnarray}
 && (ig)^2 \int_0^1 d s_1 \int_{s_1}^1 d s_2 \int_{s_2}^1 d s_3
\int_{s_3}^1 d s_4 \dot x^{\mu_1}(s_1) \dot x^{\mu_2}(s_2)
\dot x^{\mu_3}(s_3) \dot x^{\mu_4}(s_4) \nonumber \\ &&{\hskip-1cm}
\langle A_{\mu_1}^{a_1}\big( x(s_1) \big)
        A_{\mu_3}^{a_3}\big( x(s_3) \big) \rangle
\langle A_{\mu_2}^{a_2}\big( x(s_2) \big)
        A_{\mu_4}^{a_4}\big( x(s_4) \big) \rangle
\tr ( T^{a_1} T^{a_2} T^{a_3} T^{a_4}). 
\label{chamo}
\end{eqnarray} 
We will now write more explicitly one of the contributions to this multiple
integral taking into account the form of the propagator (\ref{maspropa}).
The delta function in this propagator imposes very strong restrictions on the
possible contributions. Its presence implies that the only non-vanishing
configurations are those in which the two end-points of each propagator
are at the same height. To be more concrete let us consider the
computation of (\ref{chamo}) for the trefoil knot shown in fig.
\ref{morse}. This knot is made out of four curves, $k^1$, $k^2$, $k^3$
and $k^4$, whose end-points are the four critical points $a$, $b$, $c$ and
$d$. The heights of these critical points are:
\begin{eqnarray}
 a &\rightarrow & t_1^- = t_4^-, \nonumber \\
 b &\rightarrow & t_1^+ = t_2^+, \nonumber \\
 c &\rightarrow & t_2^- = t_3^-,  \\
 d &\rightarrow & t_3^+ = t_4^+. \nonumber 
\label{abcd}
\end{eqnarray}
They are depicted in fig. \ref{morse}.

To obtain the contributions it is convenient to divide  in four parts the
circle that represents the knot in fig. \ref{morse} and join these
parts by lines representing the propagators, taking into account the
ordering of the four points to which they are attached. This ordering and
the delta function in the height imply that no line can have its two
end-points  attached to the same part. They also imply that there are no
contributions in which two end-points of different lines are attached to
one part and the other two to another part.  The contributions  are easily
depicted on the knot itself. One has been pictured in fig. \ref{morse}. For
each contribution, one must compute a sign, which takes into account the
direction in which one travels along the knot in the new parametrization.  To
be more explicit, let us write, for example, the integral associated to the
 contribution shown in fig. \ref{morse}. It takes the form:
\begin{eqnarray}
 && (ig)^2 {1\over (2\pi i)^2} {1\over {2^2}}
\int_{t_2^-<s_1<s_2<t_1^+} d s_1 d s_2 \nonumber\\
&&{\hskip-1.7cm}\Bigg({\dot z_1(s_1)-\dot z_2'(s_1) \over z_1(s_1) -
z_2'(s_1)} +{\dot z_1'(s_1)-\dot z_2(s_1) \over z_1'(s_1) -
z_2(s_1)}\Bigg)
\Bigg({\dot z_3(s_2)-\dot z_1'(s_2) \over z_3(s_2) - z_1'(s_2)} +{\dot
z_3'(s_2)-\dot z_1(s_2) \over z_3'(s_2) - z_1(s_2)}\Bigg),\nonumber\\
\label{greta}
\end{eqnarray}
where the primes denote the companion knot.
The data entering this
integral (\ref{greta}) are shown in fig. \ref{morse}. Notice that this
integral is not divergent if we take the limit $\varepsilon
\rightarrow 0$ before performing the integration. This feature is common
to all the contributions corresponding to the group factor under
consideration. Only the contributions related to framing are potentially
divergent. One could therefore remove in (\ref{greta}) the terms with primes
and the factor $1/2^2$. The integral to be computed takes the form:
\begin{equation}
 (ig)^2 {1\over (2\pi i)^2} \int_{t_2^-<s_1<s_2<t_1^+} d s_1 d
s_2 {\dot z_1(s_1)-\dot z_2(s_1) \over z_1(s_1) - z_2(s_1)}{\dot z_3(s_2)-\dot
z_1(s_2) \over z_3(s_2) - z_1(s_2)}.
\label{meza}
\end{equation} 
One of the two integrations can easily be performed, leading to:
\begin{equation} (ig)^2 {1\over (2\pi i)^2} \int_{t_2^-<s_2<t_1^+} d s_2 \log
\Bigg({z_1(s_2)-z_2(s_2) \over z_1(t_2^-) - z_2(t^-_2)}\Bigg) {\dot
z_3(s_2)-\dot z_1(s_2) \over z_3(s_2) - z_1(s_2)}.
\label{yamoko}
\end{equation}
Notice that, as argued before, this integral is finite.
Although $z_1$ and $z_2$ get close to each other when $s_2\rightarrow t_1^+$,
the singularity in the integrand, being logarithmic, is too mild to lead to a
divergent result.

We are now in  a position to write the form of the general contribution.
Notice that the most significant fact of our previous discussion is the
presence of the delta function in the height of the propagator
(\ref{maspropa}). It implies that the only non-vanishing configurations of
the propagators are those in which their two end-points have the same height;
in other words, only contributions in which the line representing the
propagator is horizontal do not vanish. This observation allows us to
rearrange the contributions to the perturbative series expansion in the
following way. Consider all possible pairings
$\{z_i(s),z_j'(s)\}$ of curves $k^i$ and $k^j_\varepsilon$,
$i,j=1,\dots,2n$, where $2n$ is the number of extrema of the Morse knot
under consideration.  A contribution at order $m$ in perturbation theory
will involve a path-ordered integral in the heights
$s_1<\dots<s_r<\dots<s_m$ of a product of $m$ propagators:
\begin{equation}
\prod_{r=1}^m {d z_{i_r}(s_r) - d z_{j_r}'(s_r)\over z_{i_r}(s_r) -
z_{j_r}'(s_r)}.
\label{elproducto}
\end{equation} 
This product is characterized by a set of $m$ ordered
pairings, each one labelled by a pair of numbers $(i_r,j_r)$ with
$r=1,\dots,m$. We will denote an ordered pairing of $m$ propagators
generically by $P_m$. One must take into account all possible ordered
pairings, \ie, one must sum over all the possible $P_m$. The group factor
that corresponds to each ordered pairing $P_m$ is simply obtained by placing
the group generators at the end-points of the propagators and taking the
trace of the product, which results after traveling along the knot. The
resulting group factor will be denoted by $R(P_m)$.

Another important fact that must be taken into account is the presence of a
product of signs. For each pairing
$P_m=\{(i_r,j_r),r=1,\dots,m\}$ there will be a contribution from their
product. Certainly, the result will be a sign that will depend on the ordered
pairing $P_m$. We will denote such a product by $s(P_m)$:

We are now in a position to write the full expression  for the
contribution to the perturbative series expansion at order $m$. It takes
the form:
\begin{equation}
 (ig^2)^m\Big({1\over 2\pi i}\Big)^m {1\over 2^m}\sum_{P_m}
\int_{t_{P_m}^-<t_1<\dots <t_r< \dots<t_m<  t_{P_m}^+} {\hskip-1cm} s(P_m)
\prod_{r=1}^m {d z_{i_r}(t_r) - d z_{j_r}'(t_r)\over z_{i_r}(t_r) -
z_{j_r}'(t_r)} R(P_m),
\label{eltung}
\end{equation}
where $t_{P_m}^+$ and $t_{P_m}^-$ are highest and lowest
heights, which can be reached by the last and first propagators of a given
ordered pairing $P_m$. This expression corresponds to the Kontsevich integral
for framed knots as presented in \cite{tung}. If one sets to zero all the
group factors associated to diagrams with isolated chords one can disregard
the primes in (\ref{eltung}). We define $Z_m(K)$ as the resulting
contribution at order $m$ divided by the dimension of the representation:
\begin{equation}
 Z_m(K)={(ig^2)^m\over {\rm dim}\, R}\Big({1\over 2\pi i}\Big)^m {1\over
2^m}\sum_{P_m}
\int_{t_{P_m}^-<t_1<\dots <t_r< \dots<t_m<  t_{P_m}^+} {\hskip-1cm} s(P_m)
\prod_{r=1}^m {d z_{i_r}(t_r) - d z_{j_r}(t_r)\over z_{i_r}(t_r) -
z_{j_r}(t_r)} R(P_m),
\label{original}
\end{equation}
This is the expression originally obtained by Kontsevich \cite{kont}.

\begin{figure}
\centerline{\epsffile{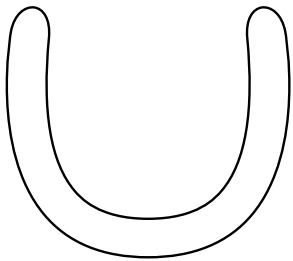}}
\caption{Unknot with four extrema.}
\label{unknot}
\end{figure}

It is well known that the Kontsevich integral is not a knot invariant. As
first pointed out by Kontsevich himself \cite{kont}, it has to be corrected 
by a subtle factor to have full invariance. Indeed, if the shape of a Morse
knot is modified in such a way that the number of extrema changes,
the Kontsevich integral (\ref{original}) is not invariant. Kontsevich proposed
the solution to this lack of invariance adding a factor now known as the
Kontsevich factor. If we denote by $U$ the unknot with the shape shown in
fig. \ref{unknot}, it turns out that the coefficients of the power expansion
of:
\begin{equation}
{1+\sum_{m=1}^\infty Z_m(K) \over (1+\sum_{m=1}^\infty Z_m(U))^{n\over 2}},
\label{verdad}
\end{equation}
where $n$ denotes the number of extrema of the knot $K$,
are truly Vassiliev invariants. The denominator of this expression is the
so-called Kontsevich factor. It is not known how to obtain this factor using
 quantum field theory arguments. This is one of the open problems from
the physical side. The study of gauge theories in non-covariant gauges
presents many problems and, as it is the case here, in many occasions they
lead to results which do not agree with their covariant-gauge counterparts.
Often the cause for the discrepancy is linked to the ambiguity in the
choice of a prescription to avoid non-physical poles. Most likely this is
not the case at least for Chern-Simons gauge theory. As  it is described in
the next section, one finds that the Kontsevich factor is needed even if no
specific prescription is taken. It is likely that the problem is related
to the fact that there is a residual gauge invariance. As we discussed above,
if one does not consider Morse knots, due to the residual gauge invariance, one
gets divergent contributions. One needs to reduce to a single point all
horizontal directions to avoid those divergences. Precisely when the number of
points on the horizontal direction changes, \ie, when the knot is deformed so
that the number of extrema gets modified one encounters the non-invariance of
the Kontsevich integral (\ref{original}). It is believed that a proper
treatment of the residual gauge invariance will lead to understand the origin
of the Kontsevich factor from a field theoretical point of view.

\vfill
\newpage

\section{Temporal gauge and combinatorial formulae}
\setcounter{equation}{0}

The perturbative studies of Chern-Simons gauge theory that we have presented
have led to two types of integral expressions for Vassiliev invariants. These
expressions are not very useful from a computational point of view. Formulae
of combinatorial type should be much preferred. There are some indications
that a general combinatorial formula for Vassiliev invariants exists.
On the one hand, the work on the study of the configuration space integrals in
the limit of flat knots \cite{alemanes} originated a combinatorial
expression for the Vassiliev invariant  of order three. On the other hand,
recent work from the mathematical side \cite{arrgpo,virgpo} supports this
conjecture. The search for the  combinatorial formula led to the
study of Chern-Simons gauge theory in the non-covariant temporal gauge
\cite{temporal}. This turns out to be the more suitable gauge to carry out
all the intermediate integrals and obtain combinatorial expressions. No other
approach has provided a combinatorial expression for the two primitive
Vassiliev invariants at order four. The temporal gauge has been also treated
in \cite{cata,vande}.

The starting point of the analysis in the temporal gauge is the same as in
the light-cone gauge. The gauge-fixing condition (\ref{light}) is the same but
now $n$ is a unit vector of the form $n^\mu=(1,0,0)$. The gauge-fixed action
and the gauge-field propagator  have also the same general form
(\ref{preochodos}) and  (\ref{nprop}). As before, the propagator presents a
pole at $np=0$, and a prescription to regulate it  is
needed. As observed in the previous sections, to construct the perturbative
series expansion  of the vev of a Wilson loop, we need the Fourier
transform  of (\ref{nprop}). We will work in the limit
$\lambda\rightarrow 0$. In the temporal gauge the momentum-space
integral that has to  be carried out has the form: 
\begin{equation} 
\Delta(x_0,x_1,x_2) = \int_M {{\rm d} ^3p\over (2\pi)^3}
{\ex^{i(p^0 x_0 +  p^1 x_1 + p^2 x_2)} \over p^0}. 
\label{espacio} 
\end{equation}  
This integral is ill-defined due to the pole at $p^0=0$. To make sense of it a
prescription has to be given to circumvent the pole. We will not take a
precise prescription. Instead, we will work in a rather general framework. Let
us first analyze the dependence of $\Delta(x_0,x_1,x_2)$ in (\ref{espacio}) on
$x_0$. The pole in $p^0$ is  avoided if, instead of (\ref{espacio}), one
analyses the derivative of $\Delta(x_0,x_1,x_2)$ with respect to $x_0$.
Considering  $\Delta(x_0,x_1,x_2)$ as a distribution one obtains: 
\begin{equation}
{\partial \Delta \over \partial x_0} = i
\delta(x_0) \delta(x_1) \delta(x_2). 
\label{lader} 
\end{equation}
Integrating this expression with respect to $x_0$, one finds that any
prescription would lead  to a result of the following form: 
\begin{equation} 
\Delta(x_0,x_1,x_2)= {i\over 2} {\hbox{\rm
sign}}(x_0) \delta(x_1)\delta(x_2) +f(x_1,x_2), 
\label{solu} 
\end{equation}
where
$f(x_1,x_2)$ is a prescription-dependent distribution. The important
consequence  of the result (\ref{solu}) is that the  dependence of
$\Delta(x_0,x_1,x_2)$ on
$x_0$ has to be in the form  ${\hbox{\rm sign}}(x_0) 
\delta(x_1)\delta(x_2)$.  This observation will be crucial in our analysis.
We will actually work with the rather general formula (\ref{solu}) for
$\Delta(x_0,x_1,x_2)$. This form  of
the propagator will allow us to introduce the notion of kernels of a Vassiliev
invariants and to design a procedure to compute combinatorial expressions for
these invariants.

As for the issue of the framing and the shift, a similar analysis as the one
in the light-cone gauge holds in this case. In the situation under
consideration, however, it is more useful to consider a specific framing, the
one in which the twist is zero or vertical framing. The linking number of a
framed knot can be expressed as the sum of the writhe plus the twist
associated to its regular projection. The writhe is the sum of the signatures
of the crossings and the twist is the number of times that the framed knot,
seen as a band, twists around its middle axis. Choosing a framing so that the
twist is zero implies that the linking number coincides with the writhe. Thus
a dependence on the writhe factorizes. This information turns out to be very
useful in the analysis of the perturbative series.

 We will  review the salient facts of the analysis of the perturbative series
expansion of the vev of a Wilson loop in the temporal
gauge. the complete analysis can be found in \cite{temporal}. Given a knot
$K$ and one of its regular knot projections,
${\cal K}$, on the $x_1,x_2$-plane which is a Morse knot in the $x_1$ and
$x_2$ directions, the perturbative series expansion for the
vev of the corresponding Wilson loop has the form:
\begin{equation}
\langle W_K^R \rangle = \langle W_{\k}^R\rangle_{{\rm temp}}\,
 \langle W_U^R\rangle^{b(\k)}, 
\label{global} 
\end{equation}
being,
\begin{equation}
{1 \over {\rm dim}\, R}\langle W_K^R \rangle = 1 + \sum_{i=1}^{\infty} v_i(K)
x^i,
\label{expansiona}
\end{equation}
and,
\begin{equation}
{1 \over {\rm dim}\, R} \langle W_{\k}^R \rangle_{{\rm temp}} = 1 +
\sum_{i=1}^{\infty} 
\hat v_i(\k)  x^i.
\label{expansionb}
\end{equation}
In these expressions $x$ denotes the coupling
constant (\ref{laequis}). The function $b({\cal K})$ is
the exponent of the Kontsevich factor, which has been conjectured to be
\cite{temporal},
\begin{equation}
b(\k) = {1\over 12} (n_{x_1}+n_{x_2}),
\label{hipotesis}
\end{equation}
where $n_{x_1}$ and $n_{x_2}$ are the critical points of the regular
projection ${\cal K}$ in both, the $x_1$ and the $x_2$ directions. In
(\ref{global}) $U$ denotes the unknot and  $\langle W_{\k}^R \rangle_{{\rm
temp}}$ is the vev of the Wilson loop corresponding to the
regular projection ${\cal K}$ as computed perturbatively in the temporal
gauge with the standard Feynman rules of the theory. Notice that though each
of the factors on the right hand side of (\ref{global}) depends on the
regular projection chosen, the left hand side does not. While the
coefficients $v_i(K)$ of the series (\ref{expansiona}) are Vassiliev
invariants the coefficients $\hat v_i(\k)$  of (\ref{expansionb}) are not. 
The latter depend on the regular projection chosen.

An explicit combinatorial form (no integrals left) of the coefficients
$\hat v_i(\k)$  in (\ref{expansionb}) would lead to a general combinatorial
formula for Vassiliev invariants. Unfortunately, this has not been obtained
yet at all orders. Only part of the contributions entering $\hat v_i(\k)$
have been explicitly written at all orders. These are the {\it kernels}
introduced in \cite{temporal}. The kernels are quantities which depend on
the knot projection chosen and therefore are not knot invariants. However,
at a given order $i$ a kernel differs from an invariant of type $i$ by
terms that vanish in signed sums of order $i$. The kernel
contains the part of a Vassiliev invariant which is the last in becoming zero
when performing signed sums, in other words, a kernel vanishes in signed sums
of order $i+1$ but does not in signed sums of order $i$. In some sense the
kernel represents the most fundamental part of a Vassiliev invariant, \ie,
the part that survives a maximum number of signed sums. Kernels plus the
structure of the perturbative series expansion seem to contain enough
information to reconstruct the full Vassiliev invariants. This was shown in
\cite{temporal} up to order four. The results obtained there will be
presented below and rewritten in a more compact form. A summary of this
approach has been presented in \cite{faroknots}. 

The expression for the kernels results after considering only the
part of the propagator (\ref{solu}) which contains the sign function.
This part involves a double delta function and therefore all the integrals
can be performed. The result is a combinatorial expression in terms of
crossing signatures after distributing propagators among all the
crossings. Of course, the contribution from this part does not depend on the
function $f(x_1,x_2)$ entering (\ref{solu}). The general expression can be
written in a universal form much in the spirit of the universal form of the
Kontsevich integral \cite{kont}. Let us consider a knot $K$ with a regular
knot projection ${\cal K}$ containing $n$ crossings. Let us choose a base
point on ${\cal K}$ and let us label the $n$ crossings by $1,2,\dots,n$ as we
pass for first time through each of them when traveling along ${\cal K}$
starting at the base point. The universal expression for the kernel
associated to
${\cal K}$ has the form:
\begin{equation}
{\cal N}(\k) = \sum_{k=0}^\infty\Bigg(
\sum_{m=1}^k \sum_{p_1,\dots,p_m =1\atop p_1+\cdots+p_m=k}^k
\sum_{i_1,\dots,i_m=1\atop i_1 < \cdots < i_m}^n
{\varepsilon_{i_1}^{p_1} \cdots \varepsilon_{i_m}^{p_m} \over
(p_1!\cdots p_m!)^2}\sum_{\sigma_{1},\dots,\sigma_{m} \atop
\sigma_{1}\in P_1,\dots,\sigma_{m}\in P_m}
{\cal T}(i_1,\sigma_{1};\dots;i_m,\sigma_{m})\Bigg).
\label{nucleos}
\end{equation}
In this equation $P_m$ denotes the permutation group of $p_m$ elements. The
factors in the inner sum, ${\cal T}(i_1,\sigma_{1};\dots;i_m,\sigma_{m})$,
are group factors which are computed in the following way:
given a set of crossings, $i_1, \dots, i_{m}$, and a set of permutations,
$\sigma_1,\dots,\sigma_m$, with $\sigma_1\in P_1,\dots,\sigma_m\in P_m$,
the corresponding group factor
${\cal T}(i_1,\sigma_{1};\dots;i_m,\sigma_{m})$ is the result of taking a
trace over the  product
of group generators which is obtained after assigning $p_1,\dots,p_m$ group
generators to the crossings $i_1, \dots, i_{m}$ respectively, and placing
each set of group generators in the order which results after traveling
along the knot starting from the base point. The first time that one
encounters a crossing
$i_j$ a product of
$p_j$ group generators is introduced; the  second
time the product is similar, but with the indices rearranged according to 
the permutation $\sigma_j\in P_j$.

The universal formula (\ref{nucleos}) for the kernels can be written in a
more useful way collecting all the coefficients multiplying a given group
factor. Recall that the group factors can be labeled by chord diagrams. At 
order $k$ one has a term for each of the inequivalent chord diagrams with $k$
chords. Denoting chord diagrams by $D$, equation (\ref{nucleos}) can be
written as:
\begin{equation} 
{\cal N}(\k) = \sum_{D} N_D(\k) D,
\label{masnucleos}
\end{equation}
where the sum extends to all inequivalent chord diagrams.
Our next task is to derive from (\ref{nucleos})  the general form of the
kernels $N_D(\k)$. The concept of kernel can be
extended to include singular knots by considering signed sums of
(\ref{masnucleos}), or, following
\cite{singular}, introducing vevs of the operators for
singular knots. If $\k^j$ denotes a regular projection of a knot $K^j$ with
$j$ simple singular crossings or double points, the corresponding universal
form for the kernel possesses an expansion similar to  (\ref{masnucleos}):
\begin{equation}
{\cal N}(\k^j) = \sum_{D} N_D(\k^j) D.
\label{sinnucleos}
\end{equation}
The general results about singular knots proved in \cite{singular} lead to
two important features for (\ref{sinnucleos}). On the one hand,
finite type implies that $N_D(\k^j)=0$ for chord diagrams $D$ with more
than $j$ chords. On the other hand, $N_D(\k^j)=2^j \delta_{D,D({\cal
K}^j)}$, where $D({\cal K}^j)$ is the chord diagram corresponding to
the singular knot projection ${\cal K}^j$. As observed above, kernels
constitute the part of a Vassiliev invariant which survives a maximum
number of signed sums.

\begin{figure}
\centerline{\epsffile{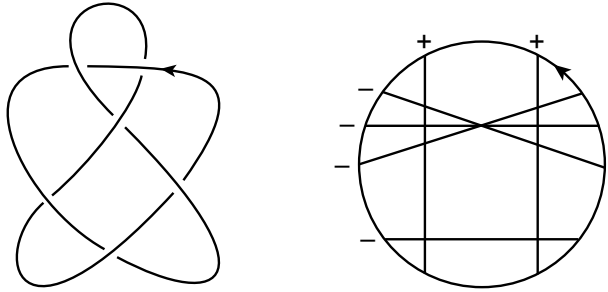}}
\caption{A regular knot projection and its corresponding Gauss diagram.}
\label{seis}
\end{figure}

    To compute $N_D(\k)$ we will introduce first the notion of the set of
labeled chord subdiagrams of a given chord diagram. We will denote this set
by $S_D$. This set is made out of a selected set of labeled chord diagrams
that we now define. 
A {\it labeled chord diagram} of order $p$ is a chord diagram with $p$
chords and a set of positive integers $k_1,k_2,\dots,k_p$, which will be
called labels, such that each chord has one of these integers attached.
The set $S_D$ is made out of labeled chord diagrams which satisfy two
conditions. These conditions are fixed by the form of the series entering
the kernels (\ref{nucleos}). We will call the elements of $S_D$ labeled
chord subdiagrams of the chord diagram $D$. They are defined as follows.
A {\it labeled chord subdiagram} of a chord diagram $D$ with $k$ chords is a
labeled chord diagram of order $p$ with labels $k_1,k_2,\dots,k_p$, $p\leq
k$, such that the following two conditions are satisfied: {\it a)}
$k_1+k_2+\cdots+k_p=k$; {\it b)} there exist  elements $\sigma_1 \in
P_{k_1},\sigma_2\in P_{k_2},\dots,\sigma_p\in P_{k_p}$ of the permutation
groups $P_{k_1},P_{k_2},\dots,P_{k_p}$ such that, after replacing the $j$-th
chord diagram by $k_j$ chords arranged according to the permutation
$\sigma_j$, for $j=1,\dots,p$, the resulting chord diagram is homeomorphic
to $D$. The number of ways that permutations  
$\sigma_1 \in P_{k_1},\sigma_2\in P_{k_2},\dots,\sigma_p\in P_{k_p}$ can be
chosen is called the multiplicity of the labeled chord subdiagram. We will
denote the multiplicity of a given labeled chord subdiagram, $s\in S_D$, by
$m_D(s)$.

The chord diagram $D$ itself can be
regarded as a labeled chord subdiagram such that its labels, or positive
integers attached to its chords, are 1. It has multiplicity 1. All
the elements of $S_D$ except
$D$ have a number of chords smaller than the number of chords of $D$. Not
all labeled chord diagrams are subdiagrams of $D$. However, given a labeled
chord diagram with labels  $k_1,k_2,\dots,k_p$ there can be different sets
of permutations leading to $D$. The number of these different sets is the
multiplicity introduced above. The elements of the sets
$S_D$ for all chord diagrams $D$ up to order four
which do not have disconnected subdiagrams are the following:
\begin{equation}
\vbox{\begin{eqnarray*}
\cdosii &\longrightarrow & \hskip0.3cm \cdosii \hbox{\hskip-0.4cm}, 
\cunodos \\
\vbox{\vskip0.9cm}
\ctresiii &\longrightarrow & \hskip0.3cm \ctresiii \hbox{\hskip-0.4cm}, 
\cdosiidosuno
\hbox{\hskip-0.4cm}, 2
\cunotres
\\
\vbox{\vskip0.9cm}
\ctresiv &\longrightarrow & \hskip0.3cm \ctresiv \hbox{\hskip-0.4cm},
 \cdosiidosuno
\hbox{\hskip-0.4cm},  \cunotres
\\
\vbox{\vskip0.9cm}
\ccuavi &\longrightarrow & \hskip0.3cm \ccuavi \hbox{\hskip-0.4cm}, 
\ctresiiidos \hbox{\hskip-0.4cm}, \cdosiitresuno
\hbox{\hskip-0.4cm},  2 \cunocuatro
\\
\vbox{\vskip0.9cm}
\ccuavii &\longrightarrow & \hskip0.3cm \ccuavii \hbox{\hskip-0.4cm}, 
 2 \cunocuatro
\\
\vbox{\vskip0.9cm}
\ccuaviii &\longrightarrow & \hskip0.3cm \ccuaviii \hbox{\hskip-0.4cm}, 
\ctresiiidos \hbox{\hskip-0.4cm}, 2\cdosiitresuno
\hbox{\hskip-0.4cm},  4 \cunocuatro
\\
\vbox{\vskip0.9cm}
\ccuaix &\longrightarrow & \hskip0.3cm \ccuaix \hbox{\hskip-0.4cm}, 
\ctresiiidosbis \hbox{\hskip-0.4cm}, 2\cdosiidosdos
\hbox{\hskip-0.4cm},   \cunocuatro
\\
\vbox{\vskip0.9cm}
\ccuax &\longrightarrow & \hskip0.3cm \ccuax \hbox{\hskip-0.4cm}, 
\ctresivdos \hbox{\hskip-0.4cm}, \ctresiiidosbis
\hbox{\hskip-0.4cm}, 2\cdosiidosdos \hbox{\hskip-0.4cm},
2\cdosiitresuno \hbox{\hskip-0.4cm}, 3\cunocuatro
\\
\vbox{\vskip0.9cm}
\ccuaxi &\longrightarrow & \hskip0.3cm \ccuaxi \hbox{\hskip-0.4cm}, 
\ctresivdos \hbox{\hskip-0.4cm}, \cdosiidosdos \hbox{\hskip-0.4cm},
\cdosiitresuno \hbox{\hskip-0.4cm}, \cunocuatro
\end{eqnarray*}}
\end{equation}
The numbers accompanying each labeled chord subdiagram denote their
multiplicity. When no number is attached to a chord of a labeled chord
diagram it should be understood that the corresponding label is 1.

In order to write our final expression for the kernels we need to recall the
notion of Gauss diagram. Given a regular projection ${\cal K}$ of a knot
$K$ we can associate to it its Gauss diagram $G({\cal K})$. The regular
projection ${\cal K}$ can be regarded as a generic immersion of a circle
into the plane enhanced by information on the crossings. The Gauss diagram
$G({\cal K})$ consists of a circle together with the preimages of each
crossing of the immersion connected by a chord. Each chord is equipped with
the sign of the signature of the corresponding crossing. An example of
Gauss diagram has been pictured in fig. \ref{seis}. Gauss diagrams are
useful because they allow to keep track of the sums involving the crossings
which enter in (\ref{nucleos}) in a very simple form. Let us consider a
chord diagram $D$ and one of its labeled chord subdiagrams $s\in S_D$. Let
us assume that $s$ has $p$ chords and labels
$k_1,k_2,\cdots,k_p$. We define the product,
\begin{equation}
\langle s, G({\cal K}) \rangle,
\label{producto}
\end{equation}
as the sum over all the embeddings of $s$ into $G({\cal K})$, each
one weighted by a factor,
\begin{equation}
{\varepsilon_1^{k_1} \varepsilon_2^{k_2}\cdots \varepsilon_p^{k_p} \over
(k_1! k_2! \cdots k_p!)^2},
\label{pesos}
\end{equation}
where $\varepsilon_1, \varepsilon_2, \dots, \varepsilon_p$ are the signatures
of the chords of $G({\cal K})$ involved in the embedding. Using
(\ref{producto})  the kernels $N_D(\k)$ entering (\ref{masnucleos}) can be
written as,
\begin{equation}
 N_D(\k) = \sum_{s\in S_D} m_D(s) \langle s, G({\cal K}) \rangle,
\label{laformula}
\end{equation}
where $m_D(s)$ denotes the multiplicity of the labeled subdiagram $s\in
S_D$ relative to the chord diagram $D$.

The product (\ref{producto}) possesses important properties. First, it is
independent of the base point chosen for the regular projection ${\cal K}$
and, correspondingly, for the Gauss diagram $G(\k)$. Second, it is of finite
type. This means that if $s$ has $j$ chords, the result of computing a
signed sum of order higher than $j$ is zero. Recall that signed sums of
order $k$ are used to define quantities associated to singular knot
projections with $k$ double points, as the ones entering
(\ref{masnucleos}). A signed sum of order $k$ contains $2^k$ terms which
correspond to the possible ways of resolving $k$ double points into 
overcrossings and undercrossings. Each one has a sign which corresponds to
the product of the signatures of the crossings involved in the $k$ double
points. If $s$ is a labeled chord diagram with $j$ chords and all its labels
take value one, the order-$j$ signed sum is $2^j$ if the configuration of
the singular projection with $j$ double points associated to such a sum
corresponds to the chord diagram $s$; otherwise its value is zero. This
fact leads to the result mentioned above stating that:
\begin{equation}
N_D(\k^j)=2^j \delta_{D,D({\cal
K}^j)},
\label{manzana}
\end{equation}
where $D({\cal K}^j)$ is the chord diagram corresponding to the
singular knot projection associated to the signed sum. Of course, the
product (\ref{producto}) vanishes if the number of chords of $s$ is bigger
than the number of chords of the Gauss diagram $G(\k)$.

The products (\ref{producto}) can be regarded as quantities of finite type
associated to Gauss diagrams $G$ whether or not these correspond to a
regular projection of a knot. Gauss diagrams can be studied as abstract
objects characterized by chord diagrams with signs assigned to their
chords. It is clear that in such a general context the quantities
$ \langle s , G \rangle $, as defined in (\ref{producto}), are of finite
type. In other words, if $s$ has $j$ chords and $G$ is an abstract Gauss
diagram, the product $\langle s , G \rangle $ vanishes under signed sums of
order higher than $j$. This observation leads to conjecture that the
product (\ref{producto}) might play an interesting role in the theory of
virtual knots \cite{virkau,virgpo}.

The terms $\langle s, G({\cal K}) \rangle$ entering (\ref{laformula}) are
related to the quantities $\chi({\cal K})$ defined in \cite{temporal}. It
is straightforward to obtain the following relations:
\begin{equation}
\vbox{\hskip-5cm
\vbox{\begin{eqnarray*}
\langle  \cunoj \hbox{\hskip-0.4cm}, G({\cal K}) \rangle &=& {1\over
(j!)^2}\chi_1({\cal K}),
\hbox{\hskip0.3cm} j \hbox{\hskip0.2cm} \hbox{\rm odd,} \nonumber \\
\vbox{\vskip0.9cm}
\langle \cdosii \hbox{\hskip-0.4cm}, G({\cal K}) \rangle
&=& 
\chi_2^A({\cal K}), \nonumber \\ 
\vbox{\vskip0.9cm}
\langle \cdosiidosdos \hbox{\hskip-0.4cm}, G({\cal K})
\rangle &=&  {1\over 16} \chi_2^C({\cal K}), \nonumber \\ 
\vbox{\vskip0.9cm}
\langle \ctresiii \hbox{\hskip-0.4cm}, G({\cal K})
\rangle &=& 
 \chi_3^B({\cal K}), \nonumber \\ 
\vbox{\vskip0.9cm}
\langle \ctresiiidos \hbox{\hskip-0.4cm}, G({\cal K})
\rangle &=&  {1\over 4} \chi_3^D({\cal K}), \nonumber \\ 
\vbox{\vskip0.9cm}
\langle \ccuaxi \hbox{\hskip-0.4cm}, G({\cal K}) \rangle
&=& 
\chi_4^A({\cal K}), \nonumber \\ 
\vbox{\vskip0.9cm}
\langle \ccuaix \hbox{\hskip-0.4cm}, G({\cal K}) \rangle
&=& 
\chi_4^C({\cal K}), \nonumber \\ 
\vbox{\vskip0.9cm}
\langle \ccuaviii \hbox{\hskip-0.4cm}, G({\cal K})
\rangle &=& 
\chi_4^E({\cal K}), \nonumber 
\vbox{\vskip0.9cm}
\end{eqnarray*}}

\vskip-10.3cm

\hskip3.2cm
\vbox{\begin{eqnarray*}
\langle \cunoj \hbox{\hskip-0.4cm}, G({\cal K}) \rangle &=& {1\over
(j!)^2}n({\cal K}),
\hbox{\hskip0.3cm} j \hbox{\hskip0.2cm} \hbox{\rm even,} \nonumber \\
\vbox{\vskip0.9cm}
\langle \cdosiidosuno \hbox{\hskip-0.4cm}, G({\cal K}) \rangle &=& 
{1\over 4} \chi_2^B({\cal K}), \nonumber \\ 
\vbox{\vskip0.9cm}
\langle \ctresiv \hbox{\hskip-0.4cm}, G({\cal K}) \rangle
&=& 
 \chi_3^A({\cal K}), \nonumber \\ 
\vbox{\vskip0.9cm}
\langle \ctresivdos \hbox{\hskip-0.4cm}, G({\cal K})
\rangle &=&  {1\over 4} \chi_3^C({\cal K}), \nonumber \\ 
\vbox{\vskip0.9cm}
\langle \ctresiiidosbis \hbox{\hskip-0.4cm}, G({\cal K})
\rangle &=&  {1\over 4} \chi_3^E({\cal K}), \nonumber \\ 
\vbox{\vskip0.9cm}
\langle \ccuax \hbox{\hskip-0.4cm}, G({\cal K}) \rangle
&=& 
\chi_4^B({\cal K}), \nonumber \\ 
\vbox{\vskip0.9cm}
\langle \ccuavi \hbox{\hskip-0.4cm}, G({\cal K}) \rangle
&=& 
\chi_4^D({\cal K}), \nonumber \\ 
\vbox{\vskip0.9cm}
\langle \ccuavii \hbox{\hskip-0.4cm}, G({\cal K}) \rangle
&=& 
\chi_4^F({\cal K}).  \nonumber 
\end{eqnarray*}}
}\label{lalista}
\end{equation}
Notice that in the second relation $n({\cal K})$ denotes the number of
crossings of the regular projection ${\cal K}$. The rest of the quantities
on the right hand side of (\ref{lalista}) were defined in \cite{temporal}.

\begin{figure}
\centerline{\hskip.4in \epsffile{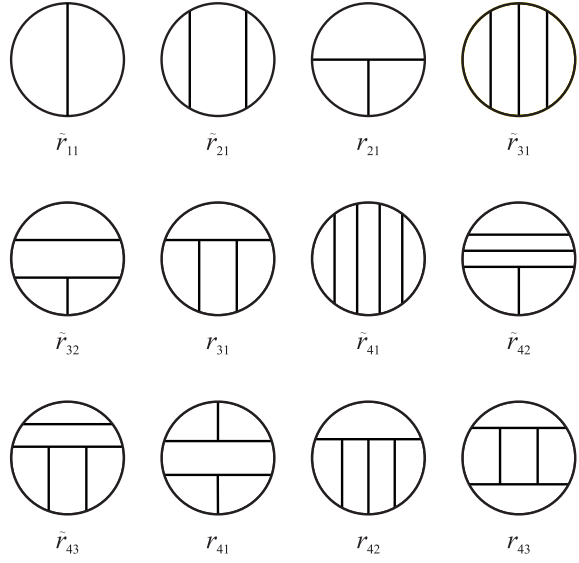}}
\caption{Choice of canonical basis up to order four which includes diagrams
with isolated chords.}
\label{canocua}
\end{figure}

\begin{figure}
\centerline{\epsffile{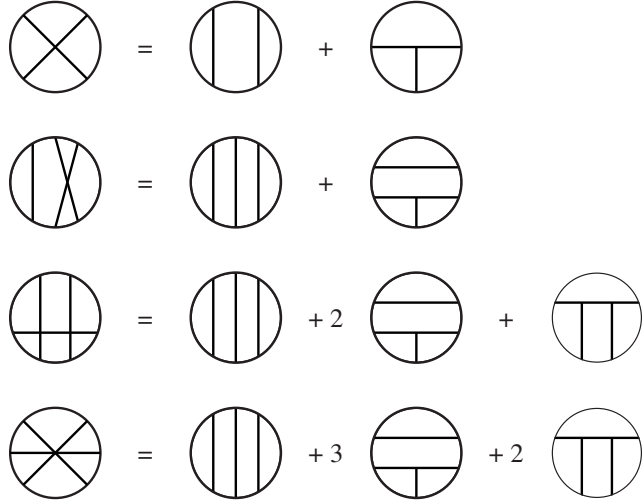}}
\caption{Expansion of chord diagrams in the canonical basis: orders two and 
three.}
\label{choruno}
\end{figure}

\begin{figure}
\centerline{\epsffile{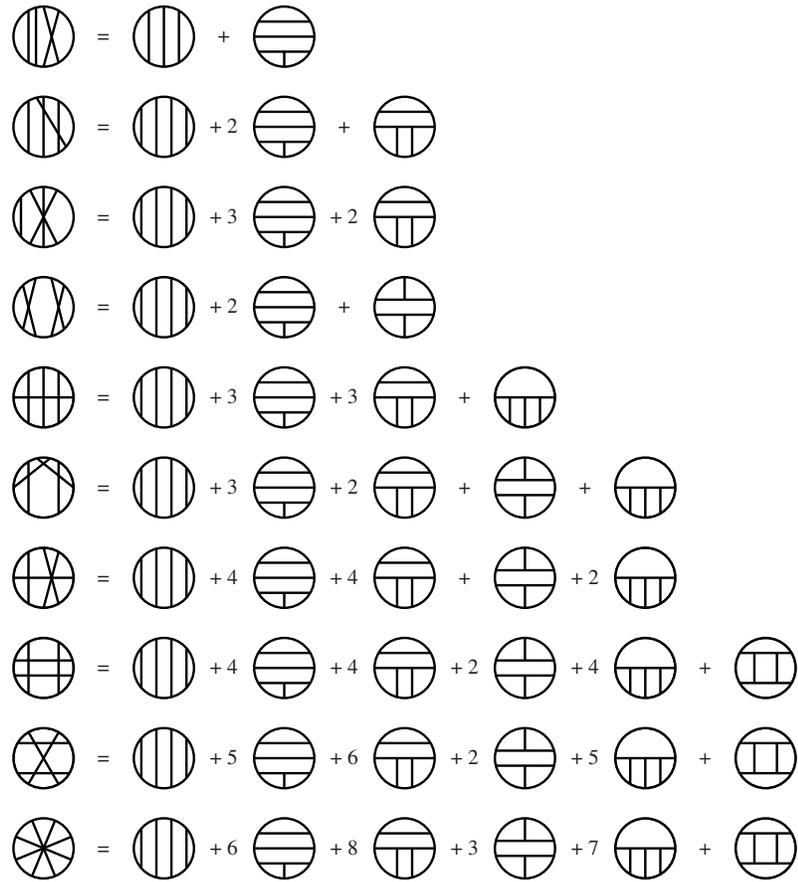}}
\caption{Expansion of chord diagrams in the canonical basis: order four.}
\label{chordos}
\end{figure}

In \cite{temporal}, Vassiliev invariants up to order four were expressed in
terms of these quantities and the crossing signatures. The strategy to obtain
them was to start with the kernels (\ref{laformula}) and exploit the properties
of the perturbative series expansion of Chern-Simons gauge theory. A special
role in the construction was played by the  factorization theorem proved in
\cite{factor}. In order to discuss some of the steps  followed in
\cite{temporal} we will describe in detail the computation of the
combinatorial expression for the Vassiliev invariant of order two. Let us begin
considering a canonical basis for the group factors where diagrams with
isolated chords are included. In fig. \ref{canocua} this basis has been
depicted up to order four. Tildes have been used to denote the diagrams with
isolated chords. These diagrams are included because 
they provide useful information when working in the vertical framing. 
Instead  of factorizing them out as in the previous sections, we will keep
them in our analysis.  In the analysis of the perturbative series
it is important to know the expressions of all the chord  diagrams
in terms of the elements of the canonical basis in fig. \ref{canocua}. These
expressions has  been collected in figs. \ref{choruno} and \ref{chordos}. 

The perturbative series expansions entering  (\ref{expansiona}) and
(\ref{expansionb}) get some modifications relative to their form in
(\ref{expansion}). We will write them  in the form: 
\begin{eqnarray}
 {1 \over {\rm dim}\, R} \langle W_{\k}^R \rangle &=& 1 + \sum_{i=1}^{\infty} 
\sum_{j=1}^{d_i}  \alpha_{ij}(K)  r_{ij}(R) x^i  + \sum_{i=1}^{\infty}
\sum_{j=1}^{\tilde d_i} \gamma_{ij}(K)  \tilde r_{ij}(R) x^i ,\nonumber \\
{1 \over {\rm dim}\, R} \langle W_{\k}^R \rangle_{{\rm temp}} &=& 1 + 
\sum_{i=1}^{\infty} 
\sum_{j=1}^{d_i} \hat \alpha_{ij}(\k)  r_{ij}(R) x^i + \sum_{i=1}^{\infty} 
\sum_{j=1}^{\tilde d_i} \hat \gamma_{ij}(\k) 
\tilde r_{ij}(R) x^i. \nonumber \\
\label{expansionc}
\end{eqnarray}
Notice that we have split the perturbative series into two sums. In the 
first sum the group factors,  and their corresponding coefficients, are
those  appearing in (\ref{expansion}), while in the second sum they are
all the  non-primitive elements coming from diagrams with isolated chords.  
The quantities $r_{ij}(R)$ and $\tilde r_{ij}(R)$ denote the respective group
factors (whose corresponding chord diagrams up to order four are depicted in
fig.
\ref{canocua}), while $d_i$ and
$\tilde d_i$ are the dimension of their basis at   order $i$. As for the
geometrical factors, 
 $\alpha_{ij}(K)$  and $\gamma_{ij}(K)$ denote the Vassiliev invariants, 
primitive or not, we are looking for, while $\hat\alpha_{ij}(\k)$  and
$\hat\gamma_{ij}(\k)$ are just the  geometrical coefficients in the 
canonical basis of the perturbative  Chern-Simons theory in the temporal
gauge.

The strategy is the following. First, the behavior of the  
unknown integrals entering  $\hat\alpha_{ij}(\k)$ and $\hat\gamma_{ij}(\k)$ is
analyzed; then  the  whole invariant is built, taking into account the
corresponding global term as dictated by (\ref{global}). Since, as shown in
\cite{factor}, the perturbative series expansion of the vev of the Wilson loop 
exponentiates  in terms of the primitive basis elements, we have the following
simple relation among primitives: 
\begin{equation}
\alpha_{ij}(K) = \hat\alpha_{ij}(\k) + b(\k)\,
\alpha_{ij}(U).
 \label{globcorrec}
\end{equation}

Let us begin with the analysis of $\hat v_i(\k)$ in (\ref{expansionb}). At 
first order we  have no correction term (recall we are using the vertical 
framing), and the temporal gauge series  provides the full regular 
invariant: 
\begin{equation}
v_1(K)= \hat v_1 ( {\cal K} ) =
\Big(\hat\gamma_{11}^E(\k)+\hat\gamma_{11}^D(\k)\Big)\, \tilde r_{11}(R)
\label{ordone}
\end{equation}
In this expression we have written the geometrical factor 
$\hat\gamma_{11}(\k)$ as a sum of two parts. The first one
$\hat\gamma_{11}^E(\k)$  is built from the part of the propagator (\ref{solu})
not containing the unknown distribution $f$. The second one,
$\hat\gamma_{11}^D(\k)$, depends on $f$. This type of decomposition can be
done in general though at higher orders is more complicated. 
The integrals made out of the $f$-dependent part of 
the  propagator (\ref{solu}) will be denoted by a superindex $D$
and a subindex which will label the chord diagram it comes from.

The general calculation requires a more
subtle labelling. Given a chord diagram, each chord in it 
represents the propagator of the theory. Our propagator (\ref{solu}) contains 
two  pieces:
the explicit one, which leads to the signatures of the crossings, and the
$f$-dependent one. The integrals arising from perturbation theory are a
sum over all the possible  ways
of identifying the chords with each of them.   So for a given   diagram  we
will  end up with different types of $D$  integrals, depending on how many
$f$-terms they contain. When  all the propagators in the  integral 
are of this kind, it will be   denoted simply by $D_{ij}$. If  only one
chord  stands for the signature-dependent part, its evaluation will simply
result in a crossing sign, $\varepsilon_m$, plus a restriction of the original
integration domain. The chord standing for this factor is
attached to the $m^{\rm th}$ crossing, which means that  the
ordered integration domain of the other chords of the $D$ integral is now
limited by the position of that crossing. The resulting
integral is written as: 
\begin{equation}  
\varepsilon_m \hat\alpha^{D_m}_{ij} \, , 
\label{noma}
\end{equation} 
with the subindex of $D$ denoting that one of the chords in the  diagram 
is attached to the $m^{\rm th}$ crossing. 

\begin{figure}
\centerline{ \epsffile{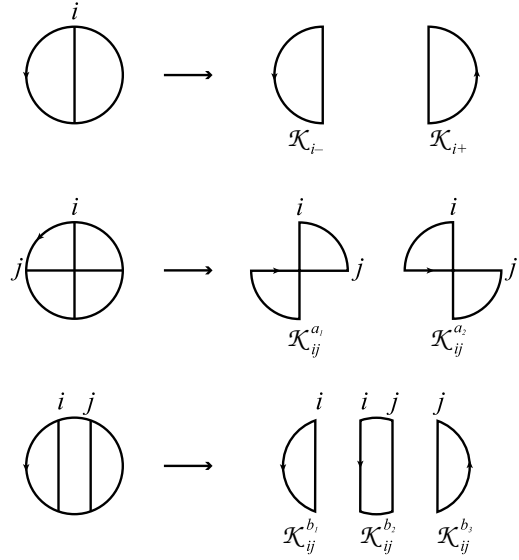}}
\caption{Splitting a knot into other knots.}
\label{subknots}
\end{figure}

More involved cases arise when the integrand contains  two 
signature-dependent terms of the propagator (\ref{solu}). In this case one
must  distinguish three subcases: both are attached to the same crossing,
both are attached to different crossings and they have the pattern of the
second case in fig. \ref{subknots}, and finally, both are attached to
different crossings but this time they have the pattern of the third
case in fig. \ref{subknots}. The set of pairs of crossings corresponding to
the second case will be denoted by ${\cal C}_a$. The one corresponding to the
third  by ${\cal C}_b$. As only invariants up to order four will be considered,
there is no need to handle the case where three or  more
signature-dependent terms of  the
propagator (but not all) are fixed to crossings. When the contribution does 
not contain $f$-dependent terms, the integral may be read from the 
kernels (\ref{nucleos}). It will be denoted by $\hat\alpha^{E}_{ij}$.

From the expression (\ref{nucleos}) for the kernels one easily
finds,  extracting the $k=1$ contribution: 
\begin{equation}
\hat\gamma_{11}^E(\k) = \sum_{i=1}^n \varepsilon_i,
\label{linka}
\end{equation}
where $n$ is the number of crossings in $\k$. This corresponds to the 
writhe, or linking number in the vertical framing, which is known to be the
correct answer for $v_1(K)$. Thus we must have 
\begin{equation}
\hat\gamma_{11}^D(\k) = 0,
\label{relat}
\end{equation}
which agrees with our general arguments, showing that contributions with an 
odd number of $f$-dependent terms vanish.

At order two, the series expansion of (\ref{expansionb}) can be expressed 
as:
\begin{equation}
\hat v_2 (\k) =  \Big(\hat\gamma_{21}^E(\k) + \hat\gamma_{21}^D(\k)\Big)\,
\tilde r_{21}(R) + \Big(\hat\beta_{21}^E(\k) + \hat\beta_{21}^D(\k)\Big)\, 
s_{21}(R),
\label{ordtwoa}
\end{equation}
where $s_{21}(R)$ is the group factor corresponding to the diagram on the
left-hand side of the first row in fig. \ref{choruno}. We will denote
by $s_{ij}$ group diagrams which appear at intermediate steps but do not
belong to the chosen canonical basis of fig. \ref{canocua}. Their associated
geometric factors will be denoted by $\beta_{ij}$. Notice that in
(\ref{ordtwoa}) we have not included terms of the form
$\sum\limits_{m=1}^n
\varepsilon_m \alpha_{ij}^{D_m}$, since they have an odd number of
$f$-dependent  terms, and should not contribute. Terms of even order in the
perturbative series expansion are invariant under a space reflection, while
terms of odd order change sign. This implies that at even (odd) orders there
are only contributions with an even (odd) number of $f$-dependent terms. In
terms of the group factors of the chosen canonical basis (see fig.
\ref{canocua}),  the last expression, after using the first equation in fig.
\ref{choruno}, takes the form: 
\begin{eqnarray}
\hat v_2 (\k) &=& \hat\gamma_{21}(\k)\, \tilde r_{21}(R) +
\hat\alpha_{21} (\k) \, r_{21}(R) \nonumber \\ 
&=&\Big( \hat\gamma_{21}^E(\k) + \hat\beta_{21}^E(\k) 
+\hat\gamma_{21}^D(\k) + \hat\beta_{21}^D(\k)\Big) \, \tilde r_{21}(R)
+\Big(\hat\beta_{21}^E(\k) +\hat\beta_{21}^D(\k)\Big)\, r_{21}(R) \nonumber \\
\label{ordtwob}
\end{eqnarray}

The computation of the two signature-dependent terms in (\ref{ordtwob}),
$\hat\gamma_{21}^E$ and $\hat\beta_{21}^E$ is easily obtained from the kernels
(\ref{nucleos}). One finds: 
\begin{equation}
\hat\gamma_{21}^E = \langle \cextra \hbox{\hskip-0.4cm} + \cunodos 
\hbox{\hskip-0.4cm}, G({\cal K}) \rangle
\label{janetuno}
\end{equation}
\begin{equation}
\hat\beta_{21}^E = \langle \cdosii \hbox{\hskip-0.4cm} + \cunodos 
\hbox{\hskip-0.4cm}, G({\cal K}) \rangle
\label{janetdos}
\end{equation}
Adding them up one obtains:
\begin{equation}
\hat\gamma_{21}^E + \hat\beta_{21}^E = {1 \over 2}\bigg( \sum_{i=1}^n
\varepsilon_i  \bigg)^2.
\label{ordtwoc}
\end{equation}
According to the factorization theorem \cite{factor} this is the whole
non-primitive regular invariant of order two, $\gamma_{21}={1\over 2}(\sum
\varepsilon_i)^2$. Thus, we conclude that the order-two $D$-type terms must 
satisfy:
\begin{equation}
 \hat\gamma_{21}^D + \hat\beta_{21}^D = 0.
\label{relata} 
\end{equation} 
One more relation is needed to get rid of the two known quantities 
$\hat\gamma_{21}^D$ and $ \hat\beta_{21}^D $. A new relation is obtained
taking into account (\ref{globcorrec}).  One easily finds: 
\begin{equation}
\alpha_{21}(K) = \langle \cdosii \hbox{\hskip-0.4cm} + \cunodos 
\hbox{\hskip-0.4cm}, G({\cal K}) \rangle   + \hat\beta_{21}^D (\k) + 
b({\cal K})\, \alpha_{21}(U),
\label{primtwoa}
\end{equation} 
where $\alpha_{21}(U)$ stands for the value of this invariant for the 
unknot. The function $b({\cal  K})$ is the unknown exponent in the global
factor in (\ref{global}). Using the fact that $\hat\beta_{21}^D (\k)$ and 
$b(\k)$  are equal  in $\k$  and $\alpha(\k)$, being $\alpha(\k)$ the
ascending diagram of  $\k$, and that the  latter is equivalent under ambient 
isotopy to the  unknot, one finds: 
\begin{equation}
\hat\beta_{21}^D (\k) = \alpha_{21}(U)\, [ 1 - b(\k) ] -
\langle \cdosii \hbox{\hskip-0.4cm} + \cunodos 
\hbox{\hskip-0.4cm}, G(\alpha({\cal K})) \rangle.
\label{primtwob}
\end{equation} 
The final expression for the invariant is:
\begin{equation}
\alpha_{21}(K) = \alpha_{21}(U) +  \langle \cdosii \hbox{\hskip-0.4cm},
G({\cal K}) \rangle - \langle \cdosii \hbox{\hskip-0.4cm},
G(\alpha({\cal K})) \rangle,
\label{primtwoca}
\end{equation} 
where  $\alpha_{21}(U)$ stands for the value of $\alpha_{21}$ for the unknot.
Recall that the ascending diagram $\alpha({\k})$ of a knot projection $\k$ is
defined as the diagram obtained by switching, when traveling along the knot
from a base point,  all the undercrossings to overcrossings. Ascending
diagrams enter often in the combinatorial expressions and it is convenient
introduce the following notation. A bar over a quantity
$L(\k)$ indicates that the same quantity for the ascending diagram has to
be subtracted, \ie:
\begin{equation}
\bar L(\k) = L(\k) - L(\alpha(\k))
\label{limon}
\end{equation}
where $\alpha({\cal K})$ denotes the standard ascending diagram of ${\cal K}$.
Using this notation, the final form for the only primitive Vassiliev
invariant at order two is:
\begin{equation}
\alpha_{21}(K) = \alpha_{21}(U) +  \langle \cdosii \hbox{\hskip-0.4cm}, \bar
G(\k) \rangle.
\label{primtwoc}
\end{equation}

The combinatorial expression (\ref{primtwoc})  agrees with the formulae
given in \cite{alemanes} and
\cite{lannes}.  Notice that  its dependence on $b({\cal K})$ has disappeared,
so up to this order  we  do  not get any condition on this function.  The
analysis at higher orders, however, imposes relations fro the function
$b({\cal K})$. All the resulting relations are consistent with the ansatz
(\ref{hipotesis}). It  is important to remark that the derivation of
(\ref{primtwoc}) that we have presented is much simpler than the one in the
covariant gauge obtained in 
\cite{alemanes}. This simplicity is rooted in the special features of the 
temporal gauge that permits to have the compact expression (\ref{nucleos}) for
the kernels, which are the essential building blocks of the combinatorial
expressions for Vassiliev invariants. 

The procedure followed at second order has been implemented in
\cite{temporal} for orders three and four. We will reproduce here the
resulting combinatorial expressions. At order three there is only one
primitive invariant. It takes the form:
\begin{equation}
\alpha_{31} (K) = \langle \ctresiii \hbox{\hskip-0.4cm} + \ctresiv
\hbox{\hskip-0.4cm} + 2 \cdosiidosuno
\hbox{\hskip-0.4cm},  G(\k) \rangle
 - \sum_{i=1}^n \, \varepsilon_i(\k) \Big[ \langle  \cdosii
\hbox{\hskip-0.4cm}, G(\alpha(\k)) \rangle \Big]_i.
\label{primthreeb}
\end{equation}
Several comments are in order to explain the quantities entering  this
expression. The sum is
over all crossings $i$, $i=1,\dots,n$, and $\varepsilon_i(\k)$ denotes the
corresponding signature. The square brackets $[$ $]_i$ enclosing a quantity
$L(\k)$ denote:
\begin{equation}
 \Big[  L(\k) \Big]_i = L(\k) - L(\k_{i_+}) -L(\k_{i_-}),
\label{platano}
\end{equation}
where the regular projection diagrams $\k _{i_+}$ and
$\k _{i_-}$ are the ones which result after the splitting of $\k$ at the
crossing point $i$ as shown in the first row of fig. \ref{subknots}. It is
clear from the list (\ref{lalista}) that these two invariants can be
written in terms of the products (\ref{producto}) and the crossing
signatures.

Combinatorial expressions for the two primitive invariants at order four
have been presented in \cite{temporal}. Their construction is based on the
use of the kernels (\ref{laformula}) and the properties of the perturbative
series expansion. As in the case of previous orders, these invariants are
expressed in terms of the products (\ref{producto}) and the crossing
signatures. Their form is more complicated than the ones at lower orders.
 At order four there are two primitive Vassiliev
invariants. We will make the same choice of basis as in
\cite{temporal}. The diagrams associated to them are: $r_{4,2}$ and
$r_{4,3}$ in fig. \ref{canonical}. They turn out to be:
\begin{eqnarray}
&& {\hskip -0.0cm} \alpha_{42}(K) = \alpha_{42}(U) +
\langle 7 \ccuaxi \hbox{\hskip-0.4cm} + 5 \ccuax \hbox{\hskip-0.4cm} + 4
\ccuaix \hbox{\hskip-0.4cm} + 2 \ccuaviii \hbox{\hskip-0.4cm} + \ccuavi
\hbox{\hskip-0.4cm} + \ccuavii \hbox{\hskip-0.4cm}  \nonumber \\ && 
\nonumber \\ && 
\hbox{\hskip6.18cm} + 8
\ctresivdos
\hbox{\hskip-0.4cm} + 2 \ctresiiidos \hbox{\hskip-0.4cm} + 8 \ctresiiidosbis
\hbox{\hskip-0.4cm} +{1\over 6} \cdosii \hbox{\hskip-0.4cm},
\bar G(\k) \rangle  \nonumber \\ && +
\sum_{i,j\in {\cal C}_a\atop i>j} \bar\varepsilon_{ij}(\k)\Bigg(
\Big[\langle \cdosii \hbox{\hskip-0.4cm}, G(\alpha(\k))\rangle \Big]_{ij}^a
-2
\Big[\langle \cdosii \hbox{\hskip-0.4cm}, G(\alpha(\k))\rangle\Big]_i - 2
\Big[\langle \cdosii \hbox{\hskip-0.4cm}, G(\alpha(\k))\rangle\Big]_j
\Bigg)
\nonumber \\ && +
\sum_{i,j\in {\cal C}_b\atop i>j} \bar\varepsilon_{ij}(\k)\Bigg(
\Big[\langle \cdosii \hbox{\hskip-0.4cm}, G(\alpha(\k))\rangle\Big]_{ij}^b -
\Big[\langle \cdosii \hbox{\hskip-0.4cm}, G(\alpha(\k))\rangle\Big]_i - 
\Big[\langle \cdosii \hbox{\hskip-0.4cm}, G(\alpha(\k))\rangle\Big]_j
\Bigg),  \nonumber \\ 
\label{primfourm}
\end{eqnarray}
and,
\begin{eqnarray}
&& {\hskip -0.1cm} \alpha_{43}(K) = \alpha_{43}(U) + 
\langle  \ccuaxi \hbox{\hskip-0.4cm} + \ccuax \hbox{\hskip-0.4cm} + 
\ccuaix \hbox{\hskip-0.4cm} + 2 \ctresiiidosbis \hbox{\hskip-0.4cm}
- {1\over 6} \cdosii \hbox{\hskip-0.4cm}, \bar G(\k) \rangle
\nonumber \\ && +
\sum_{i,j\in {\cal C}_a\atop i>j} \bar\varepsilon_{ij}(\k)\Bigg(
\Big[\langle \cdosii \hbox{\hskip-0.4cm},
G(\alpha(\k))\rangle\Big]_{ij}^a-\Big[\langle \cdosii \hbox{\hskip-0.4cm},
G(\alpha(\k))\rangle\Big]_i - 
\Big[\langle \cdosii \hbox{\hskip-0.4cm}, G(\alpha(\k))\rangle\Big]_j
\Bigg). \nonumber \\
\label{primfourn}
\end{eqnarray}
In these expressions the explicit dependence on the signatures appears in
the quantities $\bar\varepsilon_{ij}(\k)$ which are:
\begin{equation}
\bar\varepsilon_{ij}(\k) = \varepsilon_{ij}(\k) - \varepsilon_{ij}(\alpha(\k))=
\varepsilon_{i}(\k)\varepsilon_{j}(\k) -
\varepsilon_{i}(\alpha(\k))\varepsilon_{j}(\alpha(\k)).
\label{lasepsilons}
\end{equation}
The sums in which these products are involved are over double splittings of
the knot projection $\k$ at the crossings $i$ and $j$. There are two ways of
carrying out these double splittings, depending on the configuration
associated to the crossings $i$ and $j$. These are shown in the second and
third rows of fig.
\ref{subknots}. In the first one the regular projection $\k$ is split 
into two while in the second
one it is split into three. Splittings of the first type build the set
${\cal C}_a$. The ones of the second type build ${\cal C}_b$. While only
the first one is involved in the invariant $\alpha_{43}$, both appear
in $\alpha_{42}$. The new quantities entering the sums are:
\begin{eqnarray}
\Big[L(\k)\Big]_{ij}^a &=&
L(\k) - L(\k_{ij}^{a_1}) -  L(\k_{ij}^{a_2}),
\nonumber \\
\Big[L(\k)\Big]_{ij}^b &=&
L(\k) - L(\k_{ij}^{b_1}) -  L(\k_{ij}^{b_2}) -  L(\k_{ij}^{b_3}),
\label{sandia}
\end{eqnarray}
where $\k_{ij}^{a_1},\k_{ij}^{a_2},\k_{ij}^{b_1},\k_{ij}^{b_2}$ and
$\k_{ij}^{b_3}$ are the knot projections which originate after a double
splitting of $\k$, as denoted in fig.
\ref{subknots}. As in previous orders, in the  expressions
(\ref{primfourm}) and (\ref{primfourn}), the quantities
$\alpha_{42}(U)$ and
$\alpha_{43}(U)$ correspond to the value of these invariants for the
unknot. It has been proved in \cite{temporal} that the combinatorial
expressions for $\alpha_{42}(K)$ and
$\alpha_{43}(K)$ in (\ref{primfourm}) and (\ref{primfourn}) are invariant
under Reidemeister moves of fig. \ref{reide}.

Vassiliev invariants constitute vector spaces and their normalization can
be chosen in such a way that they are integer-valued. Once their value for
the unknot has been subtracted off they can be presented in many
basis in which they are integers. We will chose here a  particular basis in
which the numerical values for the invariants up to order four are rather
simple:
\begin{eqnarray}
\nu_{2}(K) &=& {1\over 4} \tilde\alpha_{21}(K), \nonumber \\
\nu_{3}(K) &=& {1\over 8} \tilde\alpha_{31}(K), \nonumber \\
\nu_{4}^1(K) &=& {1\over 8} (\tilde\alpha_{42}(K) + \tilde\alpha_{43}(K)),
 \\
\nu_{4}^2(K) &=& {1\over 4} (\tilde\alpha_{42}(K) - 5 \tilde\alpha_{43}(K)).
\nonumber
\label{basica}
\end{eqnarray}
In these equations the tilde indicates that the value for the unknot
has been subtracted, \ie,
$\tilde\alpha_{ij}(K)=\alpha_{ij}(K)-\alpha_{ij}(U)$. In Tables 1 and 2 we
have collected the value of the Vassiliev invariants (\ref{basica}) for all
prime knots up to nine crossings. Notice that we could have chosen a basis
where all the values for the trefoil knot are 1  just redefining
$\nu_{4}^1(K)$ into $\nu_{4}^1(K)-2 \nu_{4}^2(K)$. We have no done so because
$\nu_{4}^1(K)$, as defined in (\ref{basica}), has a simple shape when 
plotted versus $\nu_2(K)$. Actually, the resulting shape
has features similar to the shape which results after plotting $\nu_3(K)$
versus $\nu_2(K)$ (see \cite{faroknots} for more comments in this respect). 
The similar behavior observed for $|\nu_3(K)|$ and $\nu_{4}^2(K)$ is expected
from their general form for torus knots. As it was shown in \cite{torusknots}
and \cite{simon}, for a torus knot characterized by two coprime integers $p$
and $q$ these invariants are the following polynomials in $p$ and $q$:
\begin{eqnarray}
\nu_2(p,q) &=& {1\over 24}  (p^2-1)(q^2-1), \nonumber \\
\nu_3(p,q) &=&  {1\over 144} (p^2-1)(q^2-1)pq, \nonumber \\
\nu_4^1(p,q) &=& {1\over 288} (p^2-1)(q^2-1)p^2q^2, \\
\nu_4^2(p,q) &=& {1\over 720} (p^2-1)(q^2-1)(2p^2q^2-3p^2-3q^2-3). \nonumber
\label{toros}
\end{eqnarray}
The explicit expression of Vassiliev invariants as polynomials in $p$ and
$q$ is known up to order six \cite{torusknots}. Of course, up to order four
their value agree with the ones computed explicitly from equations
(\ref{primfourm}) and (\ref{primfourn}), as it can be checked explicitly
from the tables collected below. The only torus knots up to nine crossings
are $3_1$, $5_1$, $7_1$, $8_{19}$ and $9_1$, whose associated coprime integers
are (3,2), (5,2), (7,2), (4,3) and (9,2), respectively.

 It would be desirable to write the invariants in such a way that
signatures and split sums do not appear. Even better would be to possess
expressions where terms involving ascending diagrams are not present. It is
not known if this is possible even  for the few orders in which
combinatorial expressions for the invariants exist. There are indications
however that in order to achieve such a goal arrow diagrams as the ones
used in \cite{arrgpo} have to be introduced. The effect of the introduction
of these diagrams is to reduce the amount of embeddings entering the
product (\ref{producto}) to a selected set. Both, the expressions and the
amount of calculation could notably simplify if this is possible.
This issue is under investigation.

\vfill
\newpage

\section{Concluding remarks}
\setcounter{equation}{0}

In this paper I have presented a brief review of the developments in the
context of Chern-Simons gauge theory since the connection between this theory
and the theory of knots and links was discovered in 1988. My presentation has
started from the basics of both, the physical and mathematical theories, away
from the chronological order. I hope to have convinced  the reader
that the interplay between the physical and mathematical approaches has been
very fruitful.

Chern-Simons gauge theory has received the attention of many theoretical
physicist and it has been studied from many different points of view. The
non-perturbative study led to discover a close connection with the theory of
knots and links, which was later further analyzed in detail. Perturbative
studies led to new insights in the theory of Vassiliev invariants. These
invariants appear in the context of Chern-Simons gauge theory as the
coefficients of the perturbative series expansion. Their structure have been
analyzed working out the perturbative series expansions for different gauge
fixings. Gauge theories are very rich in this respect since they can be
studied in different gauges, each providing a particular structure for the
coefficients of the perturbative series expansion.

The perturbative analysis of the theory in the covariant Landau gauge led to
covariant integral expressions for Vassiliev invariants. These expressions
are also known as configuration space integrals. Though important, their from
is rather complicated for explicit computations. Simpler expressions for the
coefficients appear in non-covariant gauges. In the light-cone gauge one
recovers the Kontsevich integral. In the temporal gauge one finds 
combinatorial expressions. In non-covariant gauges there  is an important
issue that has not yet been solved. In both, the light-cone and the temporal
gauges, one must introduce a factor to find agreement with results in the
covariant Landau gauge or with results from a non-perturbative point of view.
The origin of this factor, called Kontsevich factor, is not understood. There
should exist a field theory argument to justify its presence. The analysis in
the temporal gauge indicates that the origin of the factor is not due to a
bad choice of the prescription to avoid unphysical poles, a standard problem
when dealing with non-covariant gauges. It seems it is related to the presence
of a residual gauge invariance. Further work in this respect should be done.
Its solution might shed some light on general problems related to
non-covariant gauges.

The perturbative analysis in the temporal gauge leads to combinatorial
expressions for Vassiliev invariants. We have constructed an approach that
avoids integral expressions making  use of the factorization theorem. It works
very successfully up to order four and, indeed, up to now, this is the only
framework which has provided combinatorial expressions for the two primitive
Vassiliev invariants at order four.  The approach opens a variety of
investigations. Certainly, a generalization of the reconstruction procedure
from the kernels  up to order four should be constructed. This could lead to a
general combinatorial formula for Vassiliev invariants. The approach is also
well suited to obtain combinatorial expressions for Vassiliev invariants for
links, a field which has not been much explored up to now \cite{alp}. Another
context in which our approach could be also very powerful is in the study of
vevs of graphs, quantities that plays an important role
in recent developments in the canonical approach to quantum gravity
\cite{ggp}. Vassiliev invariants for graphs constitute a rather unexplored
field which could lead to new sets of important invariants.

It is known that polynomial invariants, \ie, vevs of Wilson loops, do not
classify knots. On the one hand, polynomial invariants do not distinguish knots
that are not invertible (knots which are not equivalent to the ones
obtained after reversing their orientation). On the other hand, polynomial
invariants do not detect mutant knots which are not equivalent \cite{mutant}.
The question that immediately arises is whether or not  Vassiliev invariants,
being the coefficients of the power series expansion of Chern-Simons gauge
theory, have a chance to classify knots.  Fortunately, the answer to this
question is affirmative. The polynomial invariants, or vevs of Wilson loops,
are associated to a  group and a representation
(weight system). On the other hand, Vassiliev invariants are the coefficients
of the  perturbative power series expansion of Chern-Simons gauge theory. Only
after making a particular choice of group and representation one makes contact
between this series and a   polynomial invariant. But one could
consider the series as a formal one whose group factors are just the diagrams
of the space ${\cal A}$ in (\ref{chordsp}) or (\ref{trisp}).  This space might
be bigger than the space of all weight systems and therefore 
the possibility for Vassiliev invariants to classify knots could still be open.
This seems to be the case as shown in \cite{vogel}. Thus, the set of
coefficients of the formal perturbative power series expansion  (Vassiliev
invariants) is bigger than the set of vevs for any
representation and any semi-simple gauge group (polynomial invariants). This
is a promising indication that Vassiliev invariants might classify knots.
However, simple questions as whether Vassiliev invariants ever detect
nonivertibility remains open. 

Let me just finish bringing to the attention of the reader the problem of the
dimensions of each of the elements of the graded vector space of Vassiliev
invariants (\ref{chordsp}) or (\ref{trisp}). This problem is a very important
counting problem which can be addressed from both, a diagrammatic and a
group-theoretical point of view. The values of the dimensions are known only
up to order 12 \cite{jj}. No insight on the problem has been obtained from
Chern-Simons gauge theory. Could some field-theoretical method be used to
obtain the general solution to this challenging counting problem? This is
another important open question which certainly deserves further
investigation.

\vfill
\newpage

{\large\bf Acknowledgements}

\vspace{1pc}

I would like to thank my collaborators in Chern-Simons gauge theory all along
these ten years,  M. Alvarez,  L. Alvarez-Gaum\'e, J. M. Isidro, P. M. Llatas,
M. Mari\~no, E. P\'erez and A. V. Ramallo. Their insights have
led to understand many of the issues discussed in this paper.  I 
would also like to thank the organizers of the workshop ``Trends in
Theoretical Physics II", for inviting me to deliver a lecture and for
their warm hospitality. I acknowledge funds provided by the European
Commission, which supports the collaboration network `CERN-Santiago de
Compostela-La Plata' under contract C11$^*$-CT93-0315, for making possible
the organization of the workshop. This work is supported in part by DGICYT
under grant PB96-0960.
\bigskip

\vfill
\newpage

\vglue4cm

\begin{table}[hp]
\begin{center}
\begin{tabular}{|c||c|c|c|c|c|c||c|c|c|c|}\cline{1-5} \cline{7-11}
  Knot & $\nu_2$ & $\nu_3$ & $\nu_4^1$ & $\nu_4^2$  & $\;\;\;\;\;\;$
& Knot & $\nu_2$ & $\nu_3$ & $\nu_4^1$ & $\nu_4^2$ \\
  \cline{1-5} \cline{7-11}
 $3_1$ & 1 & 1 & 3 & 1 &  & $8_5$ & $-$1 & $-$3 & 1 & $-41$ \\
\cline{1-5} \cline{7-11}
 $4_1$ & $-$1 & 0 & 2 & -3 &  & $8_6$ & $-$2 & $-$3 & 7 & -36 \\
\cline{1-5} \cline{7-11}
 $5_1$ & 3 & 5 & 25 & 11  &  & $8_7$ & 2 & $-$2 & 4 & 22 \\
\cline{1-5} \cline{7-11}
 $5_2$ & 2 & 3 & 13 & 4 &  & $8_8$ & 2 & $-$1 & 5 & 12 \\
\cline{1-5} \cline{7-11}
 $6_1$ & $-$2 & $-$1 & 7 & $-12$ &  & $8_9$ & $-$2 & 0 & 14 & -34 \\
\cline{1-5} \cline{7-11}
 $6_2$ & $-$1 & $-$1 & 3 & $-13$ &  & $8_{10}$ & 3 & $-$3 & 15 & 15 \\
\cline{1-5} \cline{7-11}
 $6_3$ & 1 & 0 & 0 & 7 &  & $8_{11}$ & $-$1 & $-$2 & 2 & -27 \\
\cline{1-5} \cline{7-11}
 $7_1$ & 6 & 14 & 98 & 46 &  & $8_{12}$ & $-$3 & 0 & 14 & $-17$ \\
\cline{1-5} \cline{7-11}
 $7_2$ & 3 & 6 & 32 & 13 &  & $8_{13}$ & 1 & $-$1 & $-1$ & 17 \\
\cline{1-5} \cline{7-11}
 $7_3$ & 5 & 11 & 73 & 25 &  & $8_{14}$ & 0 & 0 & 4 & $-16$ \\
\cline{1-5} \cline{7-11}
 $7_4$ & 4 & 8 & 50 & 8 &  & $8_{15}$ & 4 & 7 & 37 & 18 \\
\cline{1-5} \cline{7-11}
 $7_5$ & 4 & 8 & 46 & 24 &  & $8_{16}$ & 1 & $-$1 & $-1$ & 17\\
\cline{1-5} \cline{7-11}
 $7_6$ & 1 & 2 & 8 & -1 &  & $8_{17}$ & $-$1 & 0 & 6 & $-19$ \\
\cline{1-5} \cline{7-11}
 $7_7$ & $-$1 & 1 & $-$1 & 3 &  & $8_{18}$ & 1 & 0 & 4 & $-9$ \\
\cline{1-5} \cline{7-11}
 $8_1$ & $-3$ & $-3$ & 13 &$-31$ & & $8_{19}$ & 5 & 10 & 60 & 35\\
\cline{1-5} \cline{7-11}
 $8_2$ & 0 & $-$1 & 3 & 30 &  & $8_{20}$ & 2 & 2 & 8 & 6 \\
\cline{1-5} \cline{7-11}
 $8_3$ & $-4$ & 0 & 30 & $-40$ &  & $8_{21}$ & 0 & $-$1 & $-$1 & $-14$ \\
\cline{1-5} \cline{7-11}
 $8_4$ & $-3$ & 1 & 21 & -39 &  &  &  &  &  &  \\
\cline{1-5} \cline{7-11}
\end{tabular}
\caption{Primitive Vassiliev invariants up to order four
for all prime knots up to eight crossings.}
\end{center}
\label{tablauno}
\end{table}

\begin{table}[hp]
\begin{center}
\begin{tabular}{|c||c|c|c|c|c|c||c|c|c|c|}\cline{1-5} \cline{7-11}
  Knot & $\nu_2$ & $\nu_3$ & $\nu_4^1$ & $\nu_4^2$ & $\;\;\;\;\;\;$
& Knot & $\nu_2$ & $\nu_3$ & $\nu_4^1$ & $\nu_4^2$  \\
  \cline{1-5} \cline{7-11}
 $9_1$ & 10 & 30 & 270 & 130 &  & $9_{26}$ & 0 & 1& $-$5 & 2 \\
\cline{1-5} \cline{7-11}
 $9_2$ & 4 & 10 & 62 & 32 &  & $9_{27}$ & 0 & 1 & 3 & $-6$  \\
\cline{1-5} \cline{7-11}
 $9_3$ & 9 & 26 & 228 & 87 &  & $9_{28}$ & 1 & 0 & $-$2 & 3 \\
\cline{1-5} \cline{7-11}
 $9_4$ & 7 & 19 & 151 & 51 &  & $9_{29}$ & 1 & $-$2 & 2 & 11 \\
\cline{1-5} \cline{7-11}
 $9_5$ & 6 & 15 & 115 & 20 &  & $9_{30}$ & $-$1 & $-$1 & 5 & $-9$ \\
\cline{1-5} \cline{7-11}
 $9_6$ & 7 & 18 & 134 & 77 &  & $9_{31}$ & 2 & 2 & 8 & 6 \\
\cline{1-5} \cline{7-11}
 $9_7$ & 5 & 12 & 78 & 47 &  & $9_{32}$ & $-$1 &2 & $-$2 & $-11$\\
\cline{1-5} \cline{7-11}
 $9_8$ & 0 & 2 & 8 & -8 &  & $9_{33}$ & 1 & $-$1 & 3 & 1 \\
\cline{1-5} \cline{7-11}
 $9_9$ & 8 & 22 & 180 & 80 &  & $9_{34}$ & $-$1 & 0 & 2 & $-3$ \\
\cline{1-5} \cline{7-11}
 $9_{10}$ & 8 & 22 & 188 & 48 &  & $9_{35}$ & 7 & 18 & 150 & 13 \\
\cline{1-5} \cline{7-11}
 $9_{11}$ & 4 & $-$9 & 57 & 10 &  & $9_{36}$ & 3 & $-$7 & 39 & 15 \\
\cline{1-5} \cline{7-11}
 $9_{12}$ & 1 & 3 & 15 & 1 &  & $9_{37}$ & $-$3 & 1 & 13 & $-7$ \\
\cline{1-5} \cline{7-11}
 $9_{13}$ & 7 & 18 & 142 & 45 &  & $9_{38}$ & 6 & 14 & 98 & 46 \\
\cline{1-5} \cline{7-11}
 $9_{14}$ & $-$1 & 2 & $-$6 & 5 &  & $9_{39}$ & 2 & $-$4 & 24 &  $-10$ \\
\cline{1-5} \cline{7-11}
 $9_{15}$ & 2 & $-$5 & 25 & 4 &  & $9_{40}$ & $-$1 & $-$1 & 3 & $-13$  \\
\cline{1-5} \cline{7-11}
 $9_{16}$ & 6 & 14 & 94 & 62 &  & $9_{41}$ & 0 & 1 & $-$9 & 18 \\
\cline{1-5} \cline{7-11}
 $9_{17}$ & $-$2 & 0 & 6 & $-2$ &  & $9_{42}$ & $-$2 & 0 & 10 & $-18$ \\
\cline{1-5} \cline{7-11}
 $9_{18}$ & 6 & 15 & 107 & 52 &  & $9_{43}$ & 1 & 2 & 14 & $-13$ \\
\cline{1-5} \cline{7-11}
 $9_{19}$ & $-$2 & 1 & 3 & 4 &  & $9_{44}$ & 0 & 1 & $-1$ & 10 \\
\cline{1-5} \cline{7-11}
 $9_{20}$ & 2 & 4 & 20 & 6 &  & $9_{45}$ & 2 & $-$4 & 20 & 6 \\
\cline{1-5} \cline{7-11}
 $9_{21}$ & 3 & $-$6 & 36 & $-3$ &  & $9_{46}$ & $-$2 & $-$3 & 3 & $-20$ \\
\cline{1-5} \cline{7-11}
 $9_{22}$ & $-$1 & 1 & 1 & 7 &  & $9_{47}$ & $-$1 & $-$2 & $-$6 & 5  \\
\cline{1-5} \cline{7-11}
 $9_{23}$ & 5 & 11 & 69 & 41 &  & $9_{48}$ & 3 & $-$5 & 29 & $-5$\\
\cline{1-5} \cline{7-11}
 $9_{24}$ & 1& 2 & 6 & $-5$  &  & $9_{49}$ & 6 & 14 & 102 & 30 \\
\cline{1-5} \cline{7-11}
 $9_{25}$ & 0 & 1 & 11 & $-14$  &  &  & & & & \\
\cline{1-5} \cline{7-11}
\end{tabular}
\caption{Primitive Vassiliev invariants up to order four 
for all prime  knots with nine crossings.}
\end{center}
\label{jpast}
\end{table}

\end{document}